\input epsf
\epsfverbosetrue
\documentclass[useAMS,usenatbib]{mn2e}
\usepackage{epsfig}
\title[Extragalactic Background Light]
{Extragalactic Background Light: a measurement at 400~nm using dark cloud shadow
\thanks{K.M., K.L., P.V. and Ch.L. dedicate this paper to the memory 
of  Gerhard von Appen-Schnur, friend and colleague, who deceased on 13 February 2013}
\thanks{Based on observations collected at the European Organisation for Astronomical 
Research in the Southern Hemisphere, 
under ESO programmes 072.A-0208(A), 082.A-0421(A), and 086.A-0201(A)}\\
{\LARGE I. Low surface brightness spectrophotometry in the area of Lynds~1642}}
\author[K. Mattila et al.]{K. Mattila$^{1}$\thanks{E-mail: mattila@cc.helsinki.fi}, 
K. Lehtinen$^{1}$,
P. V\"ais\"anen$^{2,3}$,
G. von Appen-Schnur$^{4}$ 
and Ch. Leinert$^{5}$
\\
$^{1}$Department of Physics, University of Helsinki, P.O. Box 64, FI-00014 Helsinki, Finland\\
$^{2}$South African Astronomical Observatory, P.O. Box 9 Observatory, Cape Town, South Africa\\
$^{3}$Southern African Large Telescope, P.O. Box 9 Observatory, Cape Town, South Africa\\
$^{4}$Astronomisches Institut, Ruhr-Universit\"at Bochum,  Universit\"atsstrasse 150, D-44801 Bochum, Germany \\
$^{5}$Max-Planck-Institut f\"ur Astronomie, K\"onigstuhl 17, D-69117 Heidelberg, Germany}

\begin{document}
\newcommand{\ana}{A\&A~}  % Astronomy and Astrophysics
\newcommand{\apj}{{ApJ~}}     % Astrophysical Journal
\newcommand{\mnras}{{MNRAS~}} % Monthly Notices of Royal A.S.
\newcommand{\ea}{{et~al.~}}
\newcommand{\cgs}{$10^{-9}$\,erg~cm$^{-2}$s$^{-1}$sr$^{-1}$\AA$^{-1}$}
\newcommand{\ang}{\AA \ }

\date{Accepted ; Received ; in original form }

\pagerange{\pageref{firstpage}--\pageref{lastpage}} \pubyear{2002}

\maketitle

\label{firstpage}

\begin{abstract}

{We present the method and observations for the measurement of the Extragalactic 
Background Light (EBL) utilizing the 
shadowing effect of a dark cloud. We measure the surface brightness difference between 
the opaque cloud core and its unobscured surroundings. In the difference the large
atmospheric and Zodiacal light components are eliminated and the only remaining 
foreground component is the scattered starlight from the cloud itself.  
Although much smaller, its separation is the key problem in the method. 
For its separation we use spectroscopy. 
While the scattered starlight has the characteristic Fraunhofer lines and
400~nm discontinuity the EBL spectrum is smooth and without these features.
 Medium resolution spectrophotometry at $\lambda$ = 380 - 580~nm  was performed with 
{VLT}/FORS at ESO of the surface brightness in and around the high-galactic-latitude 
dark cloud Lynds~1642. Besides the spectrum for the core with $A_V \ga 15$~mag,
further spectra were obtained for intermediate-opacity cloud positions.  
They are used as proxy for the spectrum of the impinging 
starlight spectrum and facilitate the separation of the scattered starlight 
(cf. Paper~II, \citealt{mat17b}). Our spectra reach a
precision of $\la0.5$\,\cgs\, as required to measure 
an EBL intensity in range of $\sim$1 to a few times \cgs\,. 
Because all surface brightness components are measured using the same equipment 
the method does not require unusually high absolute calibration accuracy, 
a condition which has been a problem for some previous EBL projects.}
\end{abstract}


\section{Introduction}
The importance of the Extragalactic Background Light 
(EBL) for cosmology has long been recognized, see e.g. \citet{P67} and for 
a review \citet{longair95}. The EBL at UV, optical and near infrared wavelengths consists
of the integrated light of all unresolved  galaxies along the line of sight
plus any contributions by intergalactic gas and dust and by hypothetical 
decaying relic particles. A large fraction of the energy released in 
the Universe since the recombination epoch is expected to be contained
in the EBL. An important aspect is the balance between the UV-to-NIR 
($\lambda \approx 0.1-3~\mu$m) and the mid-to-far  infrared
($\lambda \approx 5-300~\mu$m) EBL: what is lost 
 through absorption by dust in the UV--NIR will re-appear  as emission 
in the mid-to-far IR. This aspect is strongly emphasized by the detection 
of the far infrared EBL \citep{hau,pug,juvela}. Some central,  
but still largely open astrophysical problems  to which EBL measurements
can shed new light include the formation and early evolution of galaxies and
the star formation history of the Universe. There may also be significant 
numbers of Low Surface Brightness galaxies, intergalactic star clusters 
and stars escaping detection as discrete objects but contributing to the 
cumulative EBL  \citep{Vai, Zem}. In observational cosmology the 
nature of a background brightness measurement has, in principle, an advantage 
over the number count observations. When counting galaxies, whether in~magnitude 
or redshift bins, one needs to consider many kinds of selection effects which affect the 
completeness of the sample. The measurement of the EBL is not plagued by
this particular problem. However, it is hampered by the foreground components,
much larger than the EBL itself, and their elimination or accurate evaluation 
is of key importance for the direct photometric measurement of the EBL
(for a review see \citealt{mat90}).

\subsection{Recent EBL measurements}
\citet{b1} announced 'the first detection' of the EBL at 300, 550, and 800~nm. 
In their method they used a combination of space borne ({\em HST}) and ground based 
measurements. While the total sky brightness photometry with {\em HST} was free
of atmospheric effects the contribution by the Zodiacal Light, $\ga 95\%$
of the total sky, had to be measured from the ground with another telescope. 
However,  in the Zodiacal Light measurement they neglected 
some effects of the atmospheric scattered light and were not able 
to achieve for their surface photometry calibration the absolute systematic accuracy  
of $\la 0.5\%$, required for both the {\em HST} {\em and} the ground based telescope 
(see \citealt{mat03}). 
Therefore, the claim for a detection of the EBL appeared premature. 
After reanalysis of their systematic errors \citet{b5} and \citet{b7} came to the 
conclusion that \newline
'  ... the complexity of the corrections required to do absolute surface 
    (spectro)photometry from the ground make it impossible to achieve 
    1\% accuracy in the calibration of the ZL', and  
 '...the only promising strategy ... is to perform all measurements 
   with the same instrument, so that the majority of corrections and 
   foreground subtractions can be done in a relative sense, 
   before the absolute calibration is applied.'

\citet{Matsu11} have re-analyzed the {\em Pioneer 10/11} Imaging Photopolarimeter 
sky background data which were used already by \citet{Tol} to derive an
upper limit to the EBL at 440~nm. 
They have announced a detection of the EBL at 440 and 640~nm at 1.5 to 2$\sigma$ level. 
Although free of the ZL contamination, also this method is plagued by the problem 
that the EBL is merely a small difference between two large 
quantities: the total sky brightness as seen by {\em Pioneer 10/11} 
and the Integrated Starlight which makes up $\ga 95\%$ 
of the {\em Pioneer 10/11} total sky brightness signal. 
Thus, for a detection of the EBL  a very accurate ($\la 1\%$) photometric 
calibration of the {\em Pioneer 10/11} surface photometry against the diverse 
photometric systems of star catalogues used for the ISL summation is required. 

{Recently, the Long Range Reconnaissance Imager instrument aboard NASA's {\em New Horizons}
mission acquired optical broad band (440 -- 870 nm) sky background measurements
during its cruise phase beyond Jupiter's orbit. While being free of ZL, and
much less infuenced by starlight contamination than the {\em Pioneer 10/11} photometry,
the limiting factor according to the analysis of \citet{zemcov17} was the model-dependent estimation 
of the Diffuse Galactic Light. Their 2$\sigma$ upper limit was $\sim$2 times as high as the
integrated light from galaxy counts.}

In recent years an indirect method based on the absorption of TeV gamma-ray radiation
by the intervening intergalactic radiation field has been used to probe, at first, the 
{\em mid-IR} EBL \citep{Ah}. More recently it has become possible to probe
also the {\em optical} EBL \citep{Ackermann,hess,Dom,biteau}. While this method is free of the problems  
caused by the night sky component separation it does include uncertainties
of the intrinsic spectral energy distributions of the blazars,
used as probes for the gamma-ray absorption effects.

\subsection{The present EBL measurement project}
In the present situation it appears desirable to obtain another direct photometric
measurement of the EBL with an independent method. We are using the 
dark cloud shadow method as presented by \citet{mat76} and \citet{mat90}. 
It utilizes one and the same instrument for all sky components and is virtually
free of the large foreground components, i.e. the Airglow (AGL), the Zodiacal Light (ZL), 
the Integrated Starlight (ISL),  
as well as the tropospheric scattered light which together with the AGL forms the
Atmospheric Diffuse Light (ADL). The Diffuse Galactic Light (DGL), which stands for
the starlight that has been scattered by the widely distributed interstellar dust,
is eliminated as far as it originates in front of the dark cloud. 
The main task now is to account for the surface brightness (scattered light) 
of the dark cloud. In the following we will refer to the components of 
the light of the night sky by the abbreviations given here.    

The rest of this paper is organized as follows.
The dark cloud shadow method for the EBL separation is described in 
Section 2. In Section 3 the selection of observing positions in the area 
of the high-latitude dark nebula L\,1642 is described. Long-slit intermediate 
resolution spectroscopic observations, $\lambda = 380 - 580$~nm,  were carried out using the FORS1 
and FORS2 instruments \citep{app} at the {Very Large Telescope (VLT)} of ESO
and are described in Section 4. The data reduction procedures are presented in Section 5.
In Section 6 we present the calibration methods and in Section 7 the corrections
for differential effects caused by the ZL, AGL, tropospheric scattered light, 
and straylight from stars outside the measuring aperture.
Section 8 gives the observational results. 
Preliminary results of this project have been presented in \citet {mat12}.

The Paper has three appendices containing supporting data.
Appendix A describes the details of the small corrections
for differential effects caused by the ZL and  AGL. In Appendix B we 
present measurements of the straylight aureole of a star and the straylight 
correction estimates. Appendix C describes supporting intermediate 
band optical and 200~$\mu$m far-IR surface photometry  of the L\,1642 cloud area.
 
The separation of the EBL from the scattered light of the dark nebula is
described in the accompanying Paper~II \citep{mat17b}.

\begin{figure*}
\vspace{15pt}
\includegraphics[width=105mm, angle = 1]{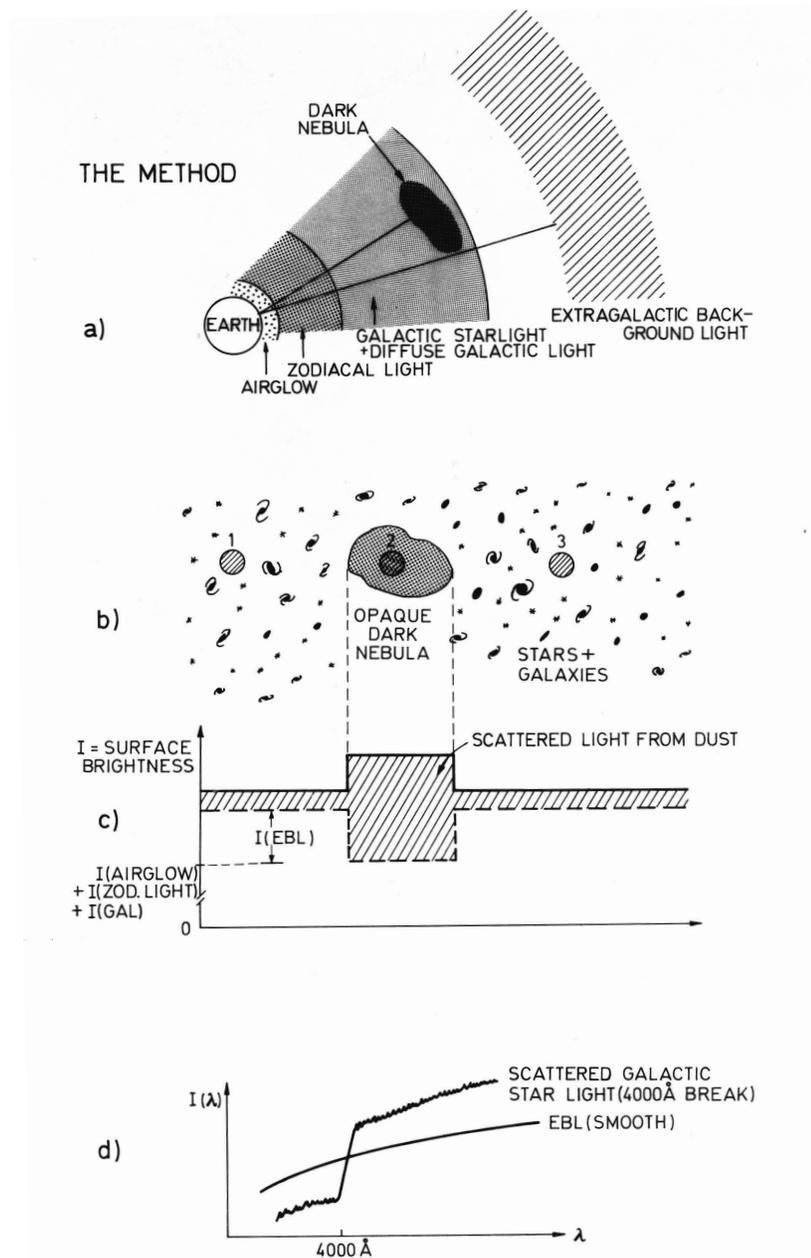}
\vspace{-5pt}
\caption{ The principle of EBL measurement with the dark cloud shadow
 method. {\bf (a)} an overview of the geometry of the night sky components involved;
 {\bf (b)} sketch of the view on the sky with three observing positions indicated:
'2' for a field in the centre of the dark cloud, '1' and '3' for fields
in transparent areas outside the dark cloud;   {\bf (c)}
If the scattered light from the interstellar dust were zero (i.e.
if the grain albedo $a = 0$), then the difference
in surface brightness between a transparent comparison area and the 
dark cloud would be due to the EBL only, and an opaque nebula
would be darker by the amount of the EBL intensity (dashed line). 
The scattered light is not zero, however.
A dark cloud in the interstellar space is exposed to the 
radiation field of the integrated Galactic starlight giving
rise to diffuse scattered light (shaded area) 
which is particularly strong in the direction of the dark cloud.
Outside the dark cloud there is still scattered light, even at high galactic latitudes,  
present as a general DGL component, though at a much lower level. 
{\bf (d)} We utilize the difference in the spectral shapes of the EBL  and the integrated 
Galactic starlight for their separation.} 
\end{figure*}

\begin{figure*}
\vspace{-20pt}
\hspace{-1.5cm}
\includegraphics[width=90mm, angle=-90]{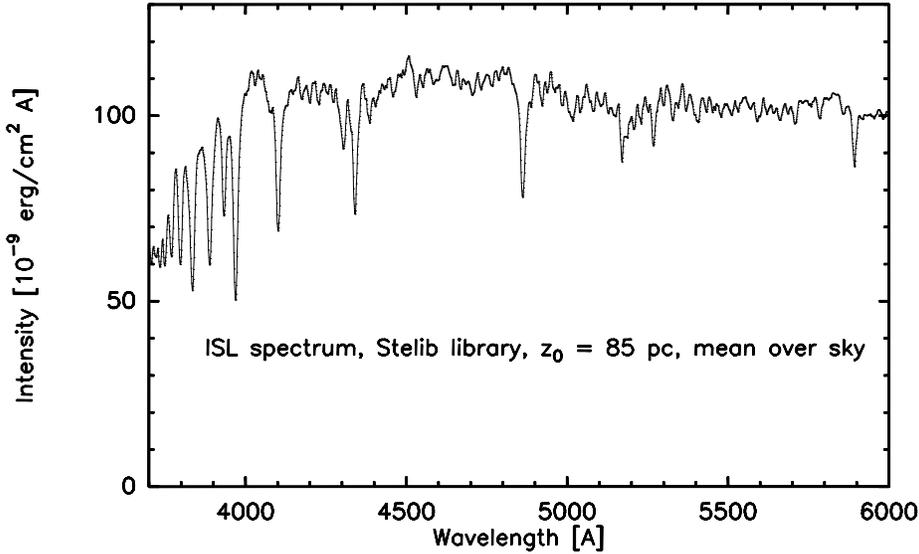}  
\vspace{-10pt}
\caption{The synthetic spectrum of integrated Galactic starlight, mean over the sky,
smoothed to our FORS resolution of 1.1~nm,
as seen by a cloud at a distance of 85~pc off the Galactic plane near the Sun. 
For the Galactic model and the stellar library, see Appendix A of Paper~II. 
Notice the several strong Fraunhofer lines and the discontinuity at 400~nm.}
\end{figure*}

 \section{The dark cloud shadow method}

The method utilizes the shadow effect of a dark cloud on the background light.
The difference of the night-sky brightness in the direction
of a high galactic latitude dark cloud and a surrounding area
that is (almost) free of obscuring dust is
due to two components only: (1) the EBL, and (2) the  starlight 
that has been diffusely scattered by interstellar dust in the 
cloud and, to a smaller extent, also by diffuse dust in its surroundings. 
Three large foreground components, i.e. the ZL, the AGL, and the 
tropospheric scattered light, are eliminated. {Fig.~1 gives an overview
of the dark cloud method.}
%
%Fig. 1a gives an overview of the geometry of the night sky components involved.
%Fig. 1b sketches the view on the sky with three observing positions indicated:
%``2'' for a field in the centre of the dark cloud, ``1'' and ``3'' for fields
%in transparent areas outside the dark cloud.   
%
%If the scattered light from the interstellar dust were zero (i.e.
%if the grain albedo $a = 0$), then the difference
%in surface brightness between a transparent comparison area and the 
%dark cloud would be due to the EBL only, and an opaque nebula
%would be darker by the amount of the EBL intensity (dashed line
%in Fig. 1c). The scattered light is not zero, however.
%A dark cloud in the interstellar space is exposed to the 
%radiation field of the integrated Galactic starlight giving
%rise to diffuse scattered light (shaded area in Fig. 1c) 
%which is particularly strong in the direction of the dark cloud.
%Outside the dark cloud there is still scattered light, even at high galactic latitudes,  
%present as a general DGL component, though at a much lower level. 

Because the intensity of the scattered light is expected to be
similar or larger than the EBL, its separation is the main task in our method.  
This can be achieved by means of spectroscopy.
%We utilize the difference in the spectra of the EBL  and the integrated Galactic starlight 
%(see simplified sketch in Fig 1d and synthetic ISL spectrum in Fig. 2). 
While the scattered Galactic starlight spectrum has
the characteristic stellar Fraunhofer lines and the  
discontinuity at 400~nm the EBL spectrum is a smooth one without 
these features {(see Fig.~1d and Fig.~2)}. 
This can be understood because the radiation from galaxies
and other luminous matter over a vast redshift range contributes
to the EBL, thus washing out any spectral lines or discontinuities.

\section{The target cloud and observed positions} 

The high galactic latitude dark nebula Lynds 1642 ($l =  210\fdg9, b = -36\fdg7$) 
has been chosen as our target. Its distance has been determined to be 
between 110 and 170~pc \citep{hearty} corresponding to a $z$-distance 
of -65 to -100~pc. It has high  obscuration ($A_{\rmn V} \ga 15$~mag)
in the centre. On the other hand, there are areas of good transparency ($A_{\rmn V} \approx 0.15$~mag) 
in its immediate surroundings within $\sim2\degr$. Its declination of -14\fdg5 
allows observations at low airmasses from Paranal. In Fig. 3 our measuring positions
for the {VLT}/FORS spectra (for which also intermediate band photometry was done) 
are shown as squares and the positions with intermediate band photometry 
only as circles (see Appendix C). The coordinates and line-of-sight 
extinction estimates for the spectroscopy positions are given in Table~1.
The extinction estimate  $A_{\rmn V}$ for position Pos8
was derived from dedicated {\em JHK} photometry of background stars using the SOFI/{NTT}
instrument at ESO/La Silla (K. Lehtinen, private communication). For the other positions 
the $A_{\rmn V}$ values were derived using {\em ISO}/ISOPHOT \citep{kessler,lemke} 
far IR (200~$\mu$m) absolute photometry, 
with Zodiacal emission and Cosmic Infrared Background subtracted, and scaled 
against 2MASS {\em JHK} near IR extinction measurements (see Appendix C for details).  

Stamps of 10x10~arcmin size centred on the  {VLT}/FORS positions, 
adopted from Digital Sky Survey blue plates, are shown in the margins:  
the high-opacity central position Pos8 with $A_V \ga 15$~mag  is shown 
in upper left, followed clockwise by the two
intermediate-opacity positions, Pos9 and Pos42 with  $A_V \approx 1$~mag, and then the 
transparent OFF-cloud positions with  $A_V \approx 0.15$~mag.

\begin{figure*}
\vspace{-160pt}   
\includegraphics[width=350mm, bb=25mm -40mm 195mm 130mm]{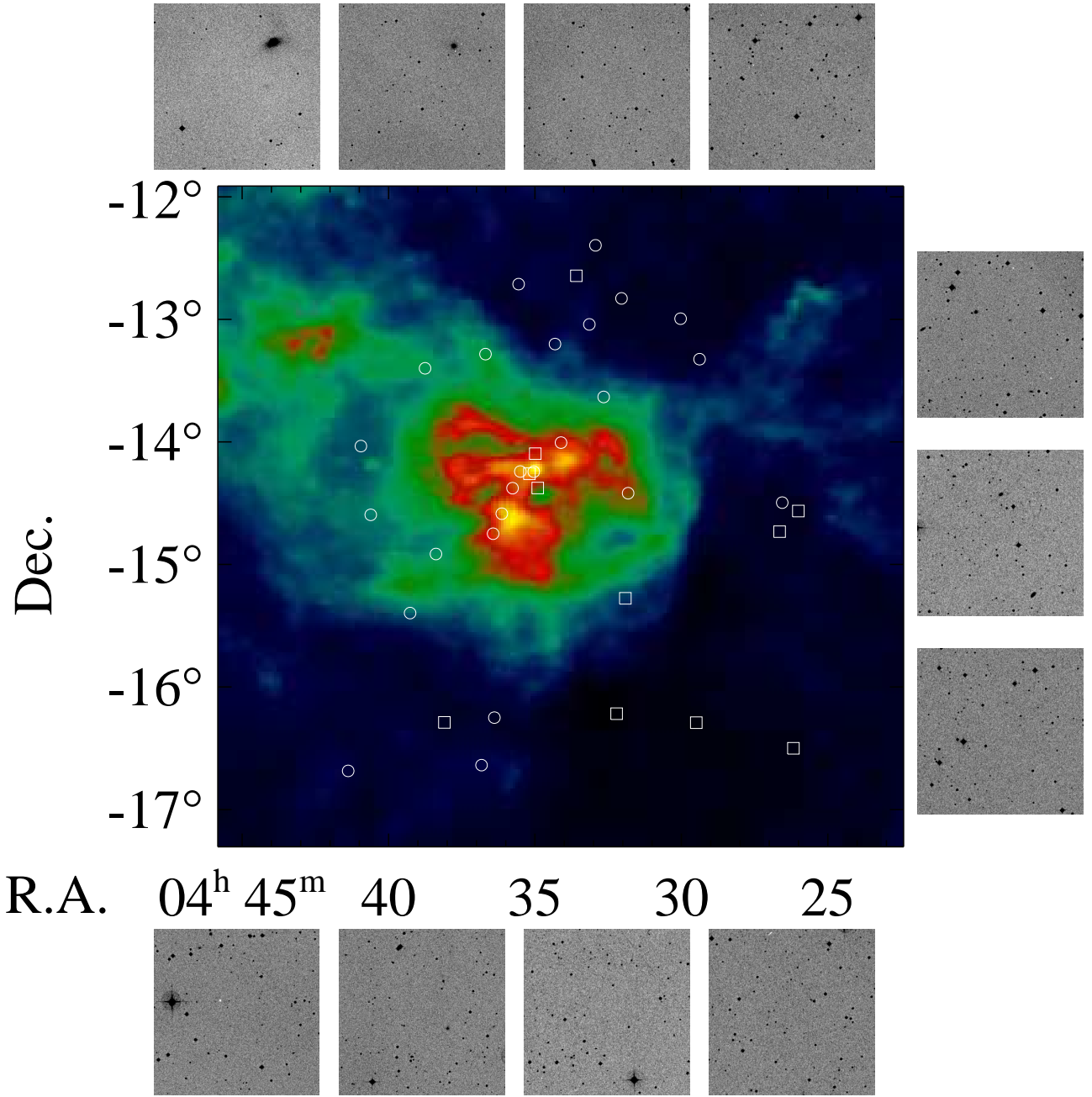}
\vspace{-350pt}
\caption{The observed positions in the L\,1642 cloud area superimposed on
a map of IRAS 100~$\mu$m emission that serves as a qualitative proxy for the interstellar 
extinction through the cloud. Positions observed spectroscopically (VLT/FORS) and photometrically
(La Silla 1-m/50-cm) are shown as squares, and positions observed only photometrically as circles.
The stamps around the map are 10 x 10~arcmin extracts from the $B$ band DSS, and are for the
positions (clockwise from upper left hand corner): Pos8, 9, 42, 13, 18, 34, 36, 32, 24, 25, and 20} 
\end{figure*}

\begin{table}
 \centering
  \caption{Data for the positions in and around L\,1642 observed in long slit 
mode with FORS. For the standard position POS08 the observed slit positions
 deviate in some cases by up to 25~arcsec from those given in the table. They are still well
within the starless dark core. See also Table 2.}
  \begin{tabular}{@{}lllcr@{}}
\hline
Name  & R.A.(J2000)   & Decl.(J2000) &I(200~$\mu$m)($\pm \sigma$) & $A_{\rmn V}$ \\
      &               &              & [MJy\,sr$^{-1}$]             & [mag] \\
\hline
POS08 &   04:35:11.2 & -14:16:26  & 58.78 (1.5)& $\ga 15$\\
POS09 &   04:34:59.5 & -14:06:36  & 23.97 (0.5)& 1.17 \\
POS42 &   04:34:54.4 & -14:23:29  & 18.14 (1.0)& 0.86 \\
\hline
POS18 &   04:33:37.0 & -12:39:25  & 5.92 (0.21)& 0.22 \\
POS20 &   04:38:04.9 & -16:18:14  & 6.49 (0.17)& 0.25 \\
POS24 &   04:29:31.0 & -16:18:12  & 3.57 (0.10)& 0.09 \\
POS25 &   04:32:13.4 & -16:13:59  & 3.71 (0.06)& 0.10 \\
POS32 &   04:26:12.3 & -16:30:21  & 3.72 (0.10)& 0.10 \\ 
POS34a&   04:26:44.2 & -14:44:16  & 4.57 (0.09)& 0.15 \\
POS34b&   04:26:54.6 & -14:44:23  & 4.57 (0.09)& 0.15 \\
POS36 &   04:26:07.7 & -14:34:32  & 4.40 (0.13)& 0.14 \\
\hline
\end{tabular}
%\end{minipage}
\end{table}

 \section{Observations}

\subsection{Observing procedure}
All our surface brightness measurements were made differentially, i.e. relative
to a 'standard position' (Pos8) in the centre of the cloud. The journal of
observations is given in Table 2. The observing procedures for 
the OFF-cloud background positions, at $\sim 2\degr$ distance from Pos8, and for the
two IN-cloud positions Pos9 and Pos42 within $\sim$10~arcmin of Pos8 were as follows:

\subsubsection{OFF-cloud positions.} 
In most cases the 'programme position PosN' measurement was bracketed by equally long
'standard position' measurements before, $I_1$(Pos8), and after, $I_2$(Pos8), 
and the surface brightness difference was calculated from:
\begin{eqnarray}
\lefteqn{\Delta I_{\rm obs}{\rm (Pos8 - PosN)} ={} } \nonumber  \\
  & & {}= \frac{1}{2}[I_1{\rm (Pos8)}+I_2{\rm (Pos8)}]-I{\rm (PosN)}
\end{eqnarray}
In two cases, because of more favourable airglow conditions, the difference was formed 
between one 'standard position' measurement bracketed by
two different 'programme position' measurements before and after (Pos18/24 on 2003-10-20
and Pos32/25 on 2004-01-25) and the surface brightness difference was calculated from:
\begin{eqnarray}
\lefteqn{\Delta I_{\rm obs}{\rm (Pos8-PosN1/N2)} ={} }   \nonumber \\
 & & {}= I{\rm (Pos8)}-\frac{1}{2}[I{\rm (PosN1)}+I{\rm (PosN2)}]
\end{eqnarray}
In these cases we obtain the mean of the surface brightness differences 
$\Delta I_{\rm obs}$(Pos8 -- PosN1) and $\Delta I_{\rm obs}$(Pos8 -- PosN2). 

For an efficient elimination of the airglow variations 
the measuring cycle had to be as short as possible. A cycle time
of 20--25 min per phase (integration + overhead time) was short enough to give
sufficiently small AGL changes in $\sim$60\% of the cases. %see next paragraph.
This still made the time spent for overheads (read-out time, telescope movements) 
a tolerable fraction (ca. 25--30\%) of the total telescope time. On the other hand,
for much shorter integration times the read-out noise would start to become
a disturbingly large fraction as compared to the photon noise, 
and the fraction of overhead time would increase.

For a time difference of  $\sim$50 min between the two 'standard position' 
measurements, 'before' and 'after', the sky surface brightness 
over the wavelength range 375--500~nm changed by $\la 5$\,\cgs\, for the best $\sim$ 30\% 
of the observations. These will be called the 'Master spectra'. 
They are listed as the first four items in Table 2. For another $\sim$ 30\% of the 
observations the change over the wavelength range 375--580~nm  was $\la$\,10\,\cgs\,; 
these are called the 'Secondary spectra'
and they are listed as the next six items in Table 2. In the case of occasional 
still larger sky level changes of  $>$\,10\,\cgs\, the data were omitted. We note that,
as described in Appendix C for the photometric measurements, the use of a parallel
monitor telescope can eliminate most of the airglow variations. An arrangement
with a spectroscopic monitor telescope was, however, not feasible for the present 
observations.

We demonstrate in Fig. 4 the different steps of our observing procedure.
The total sky spectrum for Pos24 is shown as red
while the standard position Pos8 spectra 'before' and 'after' are shown as black and
blue lines. The difference spectrum $\Delta I_{\rm obs}$(Pos8 -- {Pos24) is shown as 
blue and the difference between the two Pos8 measurements as the green line. The airglow
variability during this night was low,  $\la$\,5\,\cgs\, over the wavelength range 
375--500~nm, qualifying this observation to be included in the 'Master spectra'.    
Notice that while the difference spectrum of the two Pos8 observations still shows
substantial residuals of the strong airglow features, they have almost completely disappeared
in the  $\Delta I_{\rm obs}$(Pos8 -- Pos24) spectrum thanks to a linear variability with time of the airglow 
during this observation. The spectra displayed in Fig. 4 are calibrated but are not
corrected for extinction or other atmospheric effects, i.e. they show the spectra
as observed at the ground level. For the transformation to above-the-atmosphere see
Sections 6.3 and 7.

\subsubsection{IN-cloud positions 9 and 42.} In the case of the two IN-cloud 
positions it was possible to combine the observations of ON and OFF position
into one Observing Block with several consecutive observing pairs taken by
shifting the telescope pointing {\em along the slit}. This way the overhead time
per ON-OFF measurement pair was reduced to $\sim$12 min making an integration time
per position of 5 min a feasible choice. 
In these cases the position angle of the slit had to be adjusted 
along the direction between ON and OFF positions.   

 \begin{table*}
 \centering
 \begin{minipage}{160mm}
\vspace{2cm}
  \caption{Journal of observations. Coordinates, observing dates, integration times, 
slit position angles (from N to E) and detector chips (1, 2, or both) are given. 
The airmass $X$, airmass difference $\Delta X$, the ZL intensity  $I_{\rm ZL}$
and ZL difference  $\Delta I_{\rm ZL}$ are given in columns (7) and (8). The differences 
are in the sense Pos8 minus PosN. The ZL values are at $\lambda = 500$~nm in units of \cgs\,.}
  \begin{tabular}{@{}llllccccc@{}}

\hline
Posit- &R.A.(J2000)&Decl.(J2000)&Date and   &Slit P.A. & Obs.cycle& Airm $X$    & $I_{\rm ZL}$ & Weight\\
ion    &           &            &ESO Period &Chip(s)& Int.time &  $\Delta X$ &  $\Delta I_{\rm ZL}$& \\
(1) & (2) & (3) & (4) & (5) & (6) & (7) & (8) & (9) \\
\hline   
\multicolumn{9}{l}{Master spectra} \\
POS18&    04:33:37.0 & -12:39:25 &  2003-10-20 & 0.0             &24-8-8-18&1.073  & 102  & 1 \\
POS08&    04:35:11.2 & -14:16:26 &  72         & 1                     & 1042&-0.003& -0.4&   \\
POS24&    04:29:31.0 & -16:18:12 &                 &                      &           &  &         &   \\   
\hline 
POS24&    04:29:31.0 & -16:18:12 & 2003-10-20 & 0.0            & 8-24-8  &1.146    &103  &1   \\
POS08 &   04:35:11.2 & -14:16:26 &  72             & 1                & 1042    &0.032   &1.4&    \\
\hline 
POS24 &   04:29:31.4 & -16:18:16 & 2004-09-18 & -90.0          & 8-24-8  &1.062    &105  &1   \\
POS08 &   04:35:11.6 & -14:16:44 &   72           &  1\&2               &  1042   &0.022    &3.5 &     \\
\hline
POS20 &   04:38:04.9 & -16:18:14 & 2010-12-14 & 0.0             &8-20-8-20&1.088   &115  &0.5 \\
POS08 &   04:35:11.2 & -14:16:26 &   86            & 1\&2                 & 330   &-0.008   &3.3 &    \\
\hline
\hline
\multicolumn{9}{l}{Secondary spectra} \\
POS08 &   04:35:15.8 & -14:15:20 & 2003-11-24     &                  &            &1.025  & 121& 1 \\
POS36 &   04:26:07.7 & -14:34:32 &   72               & 0.0              & 8-36-8     &0.005  & -0.3 & \\
POS08 &   04:35:12.9 & -14:16:47 &                &  1\&2                 & 1042       &           &      &  \\
\hline
POS34b &   04:26:54.6 & -14:44:23 & 2004-01-24 & 0.0             & 8-34-8     & 1.416  & 103  & 1 \\
POS08 &   04:35:11.2 & -14:16:26 &   72            &  1\&2                &1042        &-0.034 &-0.6 &  \\
\hline
POS32 &   04:26:12.3 & -16:30:21 & 2004-01-25 & 0.0             & 32-8-8-25 &1.039   & 102 & 1  \\
POS08 &   04:35:11.2 & -14:16:26 &   72            &  1\&2                & 1042      &-0.009 &1.8&    \\
POS25  &  04:32:13.4 & -16:13:59 &                 &                 &           &            &    &    \\
\hline   
POS24 &   04:29:31.0 & -16:18:12 & 2004-02-18 & -90.0           & 8-24-8 & 1.122     & 115  &1  \\
POS08 &   04:35:11.2 & -14:16:25 &   72            &  1\&2                & 1042   & 0.004    & 1.5&   \\
\hline
POS34a &   04:26:44.2 & -14:44:16 & 2004-09-16 &  0.0            & 8-34-8 &1.053      & 107&1    \\
POS08  &  04:35:11.7 & -14:16:40 &   72            &  2                &1042    &0.017     &1.1&     \\
\hline
POS18 &   04:33:37.0 & -12:39:25 & 2011-10-02 & 0.0             &18-8-18-8-18&1.052  &104  & 0.7 \\
POS08 &   04:35:11.2 & -14:16:26 &   86            &  1\&2                &330         &-0.004&-2.7&     \\
\hline
\hline
\multicolumn{9}{l}{In-cloud positions} \\
POS42 &   04:34:54.4 & -14:23:29 & 2009-02-23 & 150.0   &$4\times(42-8)$ &1.18           &    &4    \\
POS08 &   04:35:11.2 & -14:16:26 &   82            &   1\&2       & 300            &               &    &     \\
\hline
POS42 &   04:34:54.4 & -14:23:29 & 2004-10-13 & -90.0           & 8-42-8   &1.06  &   & 1       \\ 
POS08 &   04:35:11.2 & -14:16:26 &   72            &  1\&2                &1042      &  &          &      \\
\hline 
POS42  &  04:34:54.4 & -14:23:29 & 2011-11-19 & 150.0           &$3\times(42-8)$&1.08 &  & 0.5     \\
POS08 &   04:35:11.2 & -14:16:26 &   86            &  1\&2                &315             &  &    &           \\
\hline 
POS42  &  04:34:54.4 & -14:23:29 & 2011-11-18\footnote{Moon above horizon, altitude 3-13 deg} & 150.0 &$3\times(42-8)$& 1.02 &  &0.5 \\
POS08 &   04:35:11.2 & -14:16:26 &   86            &   1\&2               & 315            &  &     &      \\
\hline
POS09  &  04:34.59.5 & -14:06:36 & 2010-11-08 & 16.1            &$4\times(9-8)$ &1.08 &   & 1     \\
POS08 &   04:35:11.2 & -14:16:26 &   86            &  1\&2                &315             &  &     &      \\
\hline
POS09 &   04:34:59.5 & -14:06:36 & 2012-12-15 & 16.1            &$4\times(9-8)$ &1.27  &   &1     \\
POS08 &   04:35:11.2 & -14:16:26 &   86            &  1\&2                & 315            &  &      &     \\
\hline
\end{tabular}
\end{minipage}
\end{table*}

\begin{figure*}
\vspace{0pt}   
\includegraphics[width=130mm, angle =-90]{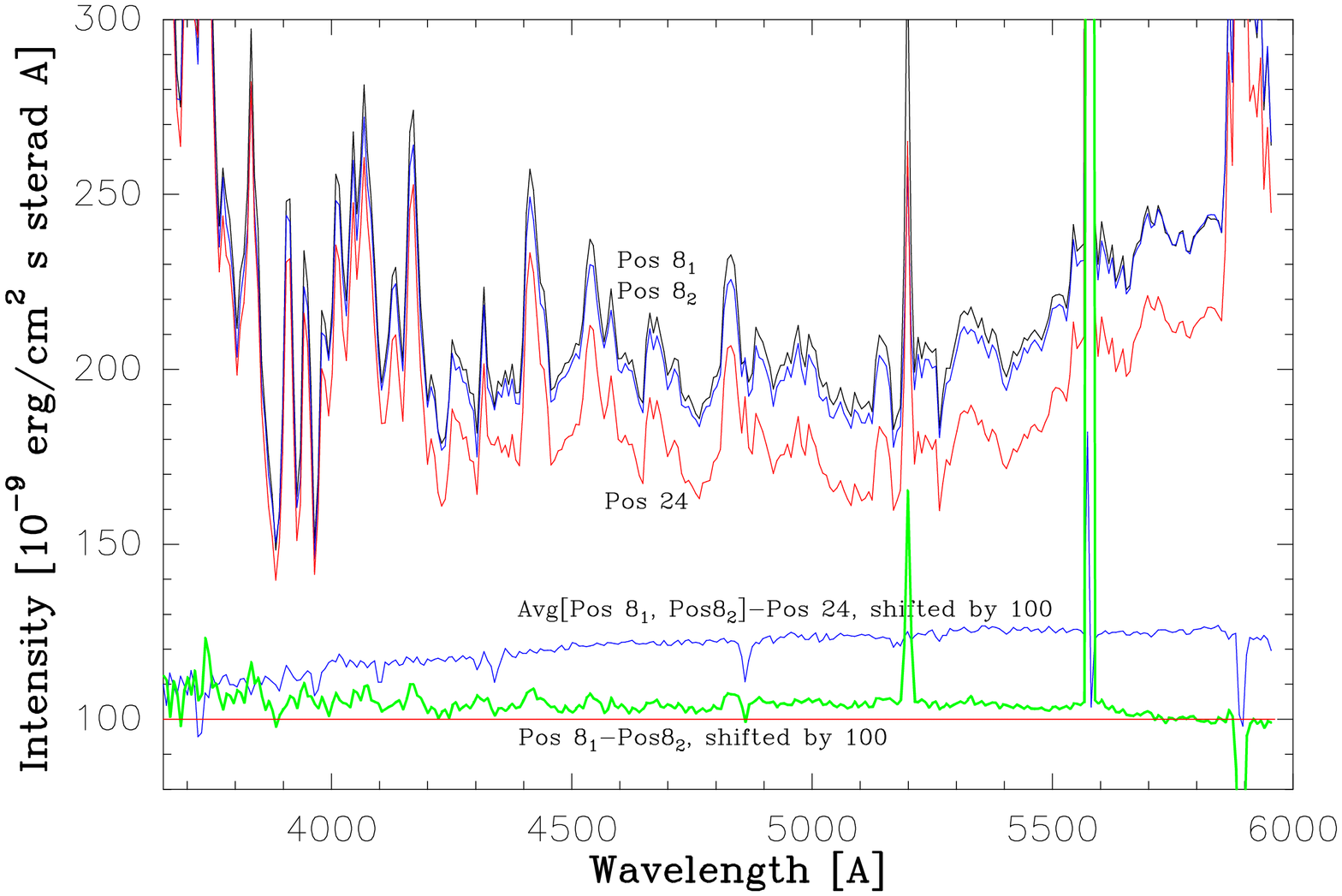}
\vspace{-20pt}
\caption{An example of the observing cycle ON-OFF-ON in the night 2004-09-18. 
The observed sky brightness spectra for the ON position (= Pos8) at the 
beginning and end of the cycle are shown as black and blue curves. Their difference,
indicating the AGL variation during the cycle, is shown as the green curve with
its zero point shifted by 100\,\cgs\,. The sky brightness spectrum for the OFF 
position (= Pos24) is shown as the red curve. The difference of the mean of the
two ON spectra and the OFF spectrum is shown as the blue curve in the bottom part
of the figure, with its zero point shifted by 100\,\cgs\,. Notice that the
AGL variability was relatively low during this night as demonstrated by the small
values of the 'quasi-continuum' difference spectrum. However, the 520.0 N{\sc i}, 
557.7 O{\sc i} and 589.0/589.6~nm NaD lines showed, as usual, a strong variability. 
The spectra are smoothed, wavelength bin is $\sim$0.7~nm and resolution $\sim$1.1~nm.
Notice that  while the difference spectrum of the two Pos8 observations (green line) still shows
substantial residuals of the strong airglow features, they have almost completely disappeared
in the Avg$[$Pos8$_1$,Pos8$_2]$--Pos24 spectrum thanks to a linear variability with time of the airglow 
during this observation. All spectra in this figure are calibrated but not corrected for
extinction and other atmospheric effects, i.e. they refer to the ground level.}  
\end{figure*}
 
\subsection{Instrumental setup}
In order to detect the Fraunhofer absorption line signature of the 
scattered light component and to separate it thereby, 
 a spectral resolution of $R \ga 400$,
or $\Delta \lambda \la 1$~nm, is needed. 
Our goal is to measure sky brightness differences (ON minus OFF)
with a precision of $\sim$0.5\,\cgs\, or $\sim$0.2\% of the
dark sky brightness, which is $\sim$(200--300)\,\cgs\, in the blue band. 
Most of the measurements were carried out in Periods 72 (2003--2004) 
and 86 (2011--2012) with FORS2 
at UT1. Some data were obtained in Period 82 with FORS1 at UT2.
 
The long slit spectrometer (LSS) mode was used with Grism 600B which gave 
a spectral coverage from 320 to 600~nm and  a nominal resolution of $R = 780$ 
or 0.075~nm\,pixel$^{-1}$ for 1-arcsec slit width. We used the 
2-arcsec slit to enable enough light to pass through and still to have 
a sufficient spectral resolution of $R \approx 400$ corresponding to 
$\sim$1~nm at 400~nm. In Periods 72 and 86 the detector was a mosaic 
of two 2k$\times$4k MIT/LL CCDs and in Period 82 a mosaic of two blue optimised 
E2V CCDs. All observations were carried out in the Service Mode.

In the setup for Period 72 and with the integration time 
of $t_{\rm int}$ = 17~min we  measured for dark sky brightness a mean count rate 
of typically  $\sim$120~e$^{-}$ and $\sim$35~e$^{-}$
per $2\times2$ read-out bin (0.15~nm $\times$ 0.252~arcsec) at $\lambda \ge 450$~nm
and  $\lambda\approx380$~nm, respectively. The detector read-out noise was 
(1.8--3.6)~e$^{-}$/bin and the dark current  1.5~e$^{-}$/bin. 
Along the 6.8~arcmin long slit we can form
an average over 1600 bins (= 400~arcsec) and in the dispersion
direction over 2~bins (= 0.3~nm) giving a total sky signal of 
 384\,$10^3$ and 112\,$10^3$ e$^{-}$ and a photon noise of
620 and 335~e$^{-}$ for $\lambda \ge 450$~nm and for $\sim380$~nm, 
respectively. The corresponding read-out noise is 100--200~e$^{-}$
while the noise caused by the dark current is negligible.
The total noise is thus $\la 650$~e$^{-}$ (corresponding to 0.4\,\cgs\,) and  
$\la$390~e$^{-}$ (corresponding to 0.9\,\cgs\,) 
 at $\lambda \ge 450$~nm and at $\sim380$~nm, respectively, 
corresponding to $\sim$0.2--0.3\% of the sky signal. 
In Periods 82 and 86 the integration time was shorter, 315--330 s, but
each position was then observed three or more times, thus giving a similar
photon noise as in the above estimate. The larger relative contribution of
the read-out noise was compensated by the reduced effect of airglow
variations thanks to the shorter cycle time. 

\section{Data reduction}

Data reduction was done using {\sc iraf} (Image Reduction and
Analysis Facility)\footnote{{\sc iraf} is distributed by the National Optical Astronomy 
Observatories, which are operated by the Association of Universities for Research
in Astronomy, Inc., under cooperative agreement with the National Science Foundation} 
\citep{Tody}. The details of the data reduction are the following.

\subsection{Bias subtraction} 
The bias signal was determined by using the overscan regions of the detector. We 
determined for each of the two halves of the detector, Chip\,1 and Chip\,2, the mean values
for each pixel column over the available $\sim$50 (Chip\,1) and  $\sim$250 (Chip\,2) rows in the 
overscan region. These mean 
values were fitted with up to 5th degree Legendre polynomials
and the resulting fitted values were subtracted from the pixel values in each column.

\subsection{Wavelength calibration and correction for geometrical distortion}
We corrected for geometrical distortion of the spectra in the
following way in {\sc iraf}: i) on a 2D~arc-lamp spectrum the emission 
features along a single dispersion line are identified with
the {\tt identify} task of {\sc iraf}; ii) the emission features at other
dispersion lines are re-identified with the {\tt reidentify} task; 
iii) the wavelengths of the identified features as a function of pixel
position are fitted with a two-dimensional function using the
{\tt fitcoords} task; iv) the geometrical correction is made with the
{\tt transform} task, after which the wavelength is a linear function 
along one axis (dispersion is constant). Despite this correction, 
subtracting the background sky emission produces artefacts 
at the edges of the brightest airglow emission
line profiles where the intensity gradient is steepest. 

\subsection{Extraction of the surface brightness signal}
In each wavelength bin of 0.15~nm width the mean value  was formed along 
the whole spatial extent of the slit, separately for the two halves of the 
detector, Chip\,1 and Chip\,2. The slit positions were generally chosen to be free 
of stars and galaxies to the limit of the DSS
images. An exception was made for the Pos 42 spectra with slit position angle 
of 150\degr: the slit was intentionally positioned to pass over a faint star
for the purpose of checking the coordinate setting accuracy and
faint-star elimination procedure. 

To eliminate the signal from faint stars and galaxies beyond the DSS limiting 
magnitude we used the {\sc idl}\footnote{http://www.exelisvis.com/ProductsServices/IDL.aspx} 
sigma-clipping method\footnote{idlastro.gsfc.nasa.gov/ftp/pro/robust/resistant\_mean.pro}.  
This method also eliminated the pixels with cosmic ray hits.
The limit above or below of which the pixels were excluded was set to 3$\sigma$
where $\sigma$ is the standard deviation of the pixel values along the slit
at a given wavelength. This standard deviation is determined mainly by the
photon statistics and depends on the wavelength and the integration time.  
Applying the 3$\sigma$ cutting meant that, for the spectra with 
an integration time of 330 sec, the pixels with surface brightness 
in the $B$ band  in excess of $\sim$140\,\cgs\, were excluded. For the  integration time 
of 1042 s the 3$\sigma$ limit was  $\sim$80\,\cgs\,.
These cut-off limits correspond to outside-the-atmosphere $B$ band surface
brightnesses of $\sim$23.3 and $\sim$23.9~mag/$\sq \arcsec$, respectively.

\section{Calibration of surface brightness measurements}

\subsection{Special aspects of surface brightness calibration}
The calibration of (spectro)photometry
of extended uniform surface brightness differs 
 from point-source photometry in the sense that 
one requires knowledge of two additional aspects: (1)
the solid angle subtended by the spectrometer or photometer 
aperture or CCD detector pixel, and (2) the aperture
correction factor.

Normally, the aperture correction factor, 
$T(A)$, is the fraction of the flux from a point source
that is contained within the aperture. The fraction $1 - T(A)$
of the point-source flux is lost outside the aperture. In the
measurement of a uniform extended source, the situation is
different: the flux that is lost from the solid angle defined by
the focal plane aperture is compensated by the flux that
is scattered and diffracted into the aperture from the sky
outside of the solid angle of the aperture. Therefore, the
intensity of an extended uniform source in \cgs\, is given by
\begin{equation}
I(\lambda) = \frac{S(\lambda) T(A)}{\Omega} C(\lambda),
\end{equation}
where  $C(\lambda)$ is the signal in instrumental 
units (count\,s$^{-1}$), $\Omega$ the solid angle of the 
aperture in steradians, and $S(\lambda)$ the sensitivity function 
in units of $10^{-9}$~erg\,cm$^{-2}$s$^{-1}$\AA$^{-1}$/(count\,s$^{-1}$)
determined from the standard-star observations.

\subsection{Observations of spectrophotometric standard stars}
As part of ESO's service mode operations
spectrophotometric standard stars were observed in the same night
(or in some cases the next night) 
with the same spectrometer setup as the surface brightnesses. However, 
instead of the 2-arcsec LSS slit the 5-arcsec MOS slit was used. 
The stars taken from ESO's list of spectrophotometric 
standards\footnote{www.eso.org/sci/observing/tools/standards/spectra/stanlis.html}: 
were in the magnitude range $V= 10.4 - 13.2$~mag.
The measurements were made using Chip\,1 of
the mosaic CCD detector only. The calibration for Chip\,2 was accomplished 
by scaling its surface brightness values to Chip\,1 values.  

A special calibration observation was carried out in the night 
2011-11-18 using the faint spectrophotometric standard star C26202 
of the {\em HST} CalSpec list\footnote{\tt www.stsci.edu/hst/crds/calspec.html} 
($V = 16.5$~mag, $B = 16.90$~mag, spectral type F8~IV).
The purpose was to check whether any detector non-linearity or
other effects appear at the faint 
signal levels. Two spectra were taken through the 
2~arcsec, slit: one unwidened spectrum and another one 
widened by drifting the star with a uniform speed along the 
slit over a distance of 100~arcsec. The latter observation produced 
in the $B$--band a signal corresponding to a surface brightness of 
22.65~mag/$\sq\arcsec$ or $\sim$240\,\cgs, similar to the night sky
brightness. The sensitivity derived from the un-widened C26202 spectrum 
was consistent with the standard star spectra 
observed in the 5~arcsec MOS slit and showed 
no colour--dependent sensitivity difference. The sensitivity ratio
derived from the widened and un-widened C26202 spectra was   unity
within $\pm$2\% between 420 and 600~nm, but rose from 420~nm towards 
ultraviolet, reaching a value of 1.17 at 370~nm. 

\subsection{Atmospheric extinction corrections}
Atmospheric extinction corrections were applied to the standard 
star flux values before using them for the calibration of the 
observed below-the-atmosphere night sky surface brightnesses. 
The average extinction 
coefficients\footnote{www.ls.eso.org/lasilla/Telescopes/2p2/D1p5M/misc/\\Extinction.html} 
as a function of wavelength for La Silla as measured
in 1974-75 by \citet{Tug} were applied. 
The average extinctions for Paranal during 2008-09 in \citet{Patat11} 
are slightly larger, by 0.02 to 0.04~mag per unit airmass between 350 
and 600~nm. However, the average $UBVRI$ extinction coefficients reported by
\citet{Patat03} for the period 03/2000 - 09/2001 are in good agreement 
with the La Silla values of  \citet{Tug}. Thus the La Silla values
 appear to be appropriate as mean extinction coefficients. They are
close to the pure Rayleigh extinction coefficients for La Silla and Paranal.

The measured  below-the-atmosphere surface brightness values were corrected 
to outside-the-atmosphere using the same extinction coefficients as 
for stellar photometry. This is justified
by the fact that we are observing surface brightness differences over an area 
of at most a few degrees. Therefore, the aerosol scattering part of extinction 
which is removing photons from the line-of-sight beam is not 
compensated for by scattering back into the beam
from the surrounding sky. This would be the case for much broader surface
brightness distributions like that of the the ZL, extending 
over several tens of degrees on the sky. 

 The extinction coefficients vary from night to night. However,
since both the standard stars and the surface brightness
target positions were observed at small airmasses ($< 1.25$) the
night-to-night variations of the extinction coefficients have only
minor influence. For example, for a deviation of the true extinction coefficient
from the mean value of 0.41~mag/airmass at 370~nm by 0.05~mag/airmass the
intensity error introduced is $< 1.2$\% if the standard star is at airmass
1.0 and the target at 1.25 or vice versa. Similar or smaller errors 
are obtained at other (longer) wavelengths. The errors of calibration stars' 
extinction correction has only a minor effect on the final results, see Section 8.2.   

\subsection{Aperture correction}
The standard star spectra were extracted from a stripe with a width of 
typically 10 pixels $(2\times2)$ or 2.5~arcsec, corresponding to the core part
of the PSF of the star image. 
In order to estimate the fraction of energy in the standard star's image 
which falls outside of this 2.5~arcsec strip in the spatial direction  
and outside of the 5~arcsec slit in the dispersion direction we have 
stacked 17 standard star observations from all nights in Period 72. 
In this stacked image the star's PSF profile can be traced out to a distance
of $\sim$25~arcsec. Assuming that the energy distribution in the star image
is centrally symmetric we have estimated the energy falling outside
of the $2.5 \arcsec \times 5 \arcsec$ extraction slot up to a distance 
of 20~arcsec, separately for three wavelength ranges,
350--425, 425--500, and 500--600~nm. No wavelength dependence was found and 
the mean value and its mean error extracted from these three wavelength 
slots was  $0.143\pm0.006$. The seeing during the standard star observations
was mostly $\le 1.5$~arcsec,  
and thus its variations did not influence the fractional energy outside
of  $2.5$~arcsec.  For the surface brightness measurements the seeing does not matter.

The part of energy falling outside of 20~arcsec was estimated in two steps:
(1) for the range $100\arcsec - 1400\arcsec$ we have made  with {VLT}-UT2/FORS1  
LSS spectrophotometric measurements of the star aureole using Sirius as light 
source (see Appendix B). With $I_{\rm stray}(r)$ the stray-radiation intensity,
 and $r$ the angular distance from the star the relationship
 log\,$I_{\rm stray}(r)$ vs. log\,$r$
was found to be closely linear with a slope of $\sim$-1.99. Thus we extrapolated
the relation up to 2\degr\, and found that within the range of 100\arcsec -- 2\degr\, 
the energy fraction was 0.038, closely the same in blue and visual.
(2) An interpolation between the two sets of measurements gave an energy
fraction of 0.029 between 20 and 100~arcsec.  With an estimated uncertainty
of $\pm50$\% for the energy fractions between 20\arcsec -- 100\arcsec\, and 
100\arcsec -- 2\degr\,, the total error was estimated to be $\sim \pm 0.05$. Summing up, we 
found that the total fraction of star flux lost was $1-T(A) = 0.21 \pm 0.05$    
and the aperture correction factor was found to be $T(A) = 0.79 \pm 0.05$.

\subsection{Width of the slit}
The width of the slit used for the surface photometry was during all three
periods $2.00$~arcsec and has an uncertainty of $\pm 0.02$~arcsec 
as given in the FORS1/2 Commissioning Report 
(non-public extensive version, M. van den Ancker, private communication 2014).
The solid angle
corresponding to one read-out pixel of $0.252$~arcsec in the spatial direction,
i.e. the area of $0.252\arcsec \times 2.00\arcsec$, 
was thus $\Omega = (1.185 \pm 0.01) \cdot 10^{-11}$~sterad.
 
\section{Differential corrections for Zodiacal Light, Atmospheric Diffuse Light and 
straylight from stars}

As explained in Section 2
the atmospheric, interplanetary and Galactic night sky components originating
closer to us than the dark cloud are eliminated; this is true as long as 
they are constant over the area covered by our target positions and the duration
of the ON--OFF--ON cycle. 
We need to consider only the differential effects caused by their time variation
and the spatial differences over $\la$2\fdg5.
For the two IN-cloud positions, Pos9 and Pos42, the angular distance to the standard
position was small enough, $\sim$10~arcmin, to, make such corrections superfluous.   

The total diffuse sky brightness observed at a transparent OFF-position (= PosN) 
at airmass $X$ is given by
\begin{equation}
I^{X}_{\rm obs}{\rm (PosN)} = I_{\rm ZL}{\rm (PosN)} e^{-\tau(\lambda) X} + I^{X}_{\rm ADL} 
\end{equation}
where $I_{\rm ZL}$ is the ZL intensity outside the atmosphere and $\tau(\lambda)$
is the extinction coefficient.  The ADL intensity, 
$I^{X}_{\rm ADL}$, is the sum the AGL and the three tropospheric scattered 
light components created by AGL, ZL and ISL as sources of light
\begin{equation}
I^{X}_{\rm ADL} = I^{X}_{\rm AGL}+ I^{X}_{\rm SCA}({\rm AGL})+ I^{X}_{\rm SCA}({\rm ZL}) 
    + I^{X}_{\rm SCA}({\rm ISL})
\end{equation} 
For the ON-position (= Pos8), observed at the slightly different airmass of  $X+\Delta X$,
the  total diffuse light intensity is
\begin{eqnarray}
\lefteqn{I^{X+\Delta X}_{\rm obs}{\rm (Pos8)} = I_{\rm ZL}{\rm (Pos8)} e^{-\tau(\lambda) (X+\Delta X)} 
+I^{X+\Delta X}_{\rm ADL} +{} } \nonumber\\
& & {}+ \Delta I^0{\rm(Pos8-PosN)} e^{-\tau(\lambda) (X+\Delta X)}
\end{eqnarray}
where  $\Delta I^0{\rm (Pos8-PosN)}$ is the (extraterrestrial) excess surface brightness of 
Pos8 relative to the transparent OFF position (PosN) outside the cloud. This is the quantity
(spectrum) we want to derive for the several OFF positions.

The observed sky brightness difference between the ON and OFF position
is thus given by:
\begin{eqnarray}
\lefteqn{\Delta I_{\rm obs}{\rm (Pos8-PosN)}=\Delta I^0{\rm (Pos8-PosN)} e^{-\tau(\lambda) (X+\Delta X)} + } \nonumber \\
\lefteqn{+ [I_{\rm ZL}{\rm (Pos8)} e^{-\tau(\lambda) (X+\Delta X)} - I_{\rm ZL}{\rm (PosN)} e^{-\tau(\lambda) X}] +} \nonumber \\
\lefteqn{+ [I^{X+\Delta X}_{\rm ADL}- I^X_{\rm ADL}] } 
\end{eqnarray}

The individual correction terms for the ZL and ADL 
are described in Appendix A. The corrections turn out to be  
$\sim$ 1 to 3\,\cgs. To some extent they have for the individual OFF
positions opposite sign and cancel out in the mean values.

\begin{figure*}
\vspace{0pt}
\hspace{-70pt}   
\includegraphics[width=130mm,angle=-90]{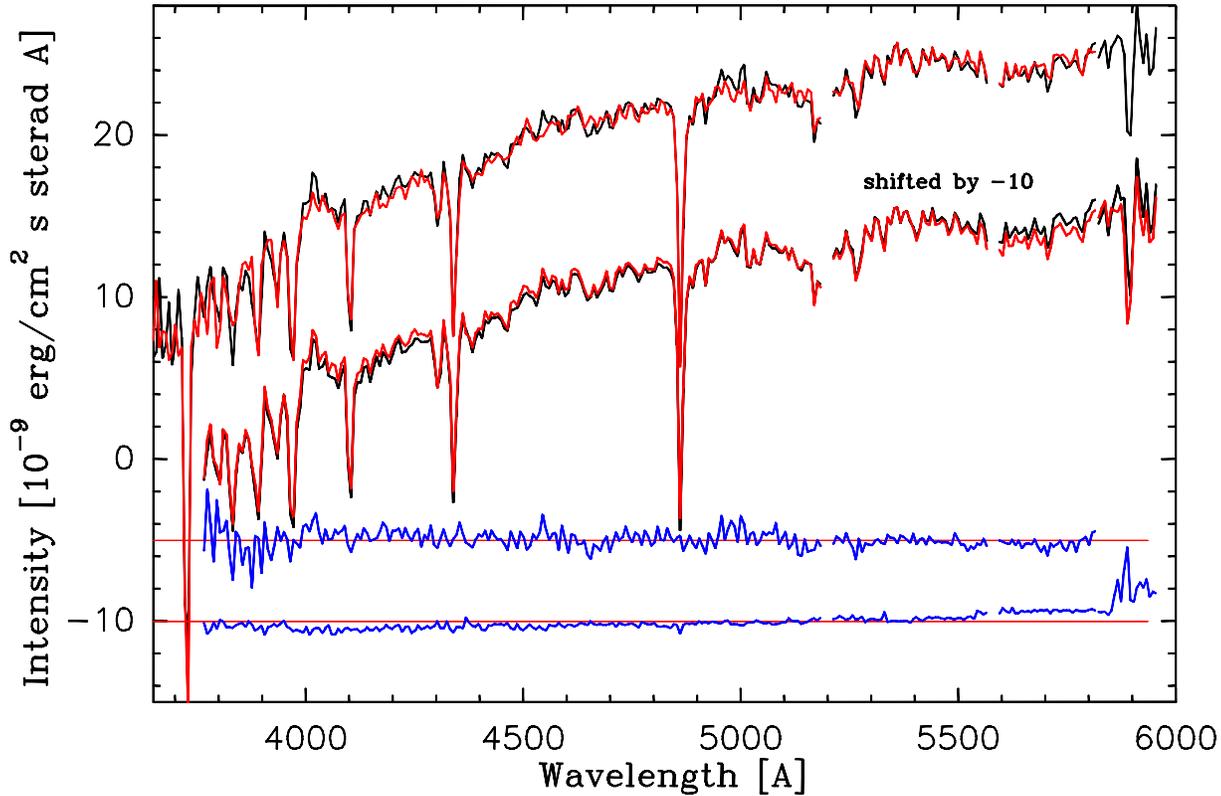}
\vspace{-20pt}
\caption{The observed surface brightness spectra for Pos8,  $\Delta I^{0}$(Pos8--OFF),  
i.e. the difference relative to the average surface brightness of several OFF positions around the cloud. 
The two {curves} 'shifted by -10 units' display  
the two 'Master mean' spectra, using either method A (black curve) or method B (red curve) for the 
atmospheric diffuse light (ADL) correction. 
Their difference is shown as the bottommost blue curve with zero 
level indicated as red line.
For the extreme red end, $\lambda = 582-595$~nm, only the results for one OFF position (Pos20) 
in the exceptionally quiescent night 2010-12-14 are included. 
The 'Secondary mean' spectrum is shown as the uppermost red curve. 
%It is seen to closely agree with 
The topmost black curve is the mean of the two 'Master' spectra,
methods A and B. The difference of the 'Master' and 'Secondary' mean spectra is shown as 
the upper blue curve, with the zero level indicated as red line.
The spectra have been smoothed with a 5-pixel boxcar function and have a bin size of  $\sim$0.7~nm
and a resolution of  $\sim$1.1~nm. The spectra have gaps at the strong airglow lines at 520.0 and 557.7~nm.  
For the purpose of demonstrating the presence of the [O{\sc ii}] 372.8~nm line and 
in order to avoid confusion the wavelength section 365--380~nm is shown for the upper two spectra only.
The spectra have been corrected for extinction and other atmospheric effects, i.e. they
refer to outside-the-atmosphere.}
\end{figure*}

The surface brightness observed towards blank areas of sky, even if
devoid of any resolved stars or galaxies, still contains instrumental 
straylight from stars that are outside of the observed area.
Depending on their brightness stars up to several tens of arcminutes away can have 
a substantial straylight contribution.
We have measured the straylight with FORS1 at UT2 over the angular offset 
range of  $r$ = 100\arcsec--1400\arcsec\,. 
The results are presented in Appendix B for several wavelength 
slots between  360 and 580~nm. 
These straylight profiles were used to calculate the $B$-- and $V$--band straylight 
intensities for each target position. 
The straylight corrections are given in Table B1; 
they are small enough to be neglected in the 
further analysis. No differential straylight corrections are needed for Pos9 and Pos42, either.

\begin{figure*}
\vspace{0pt}   
\includegraphics[width=130mm, angle=-90]{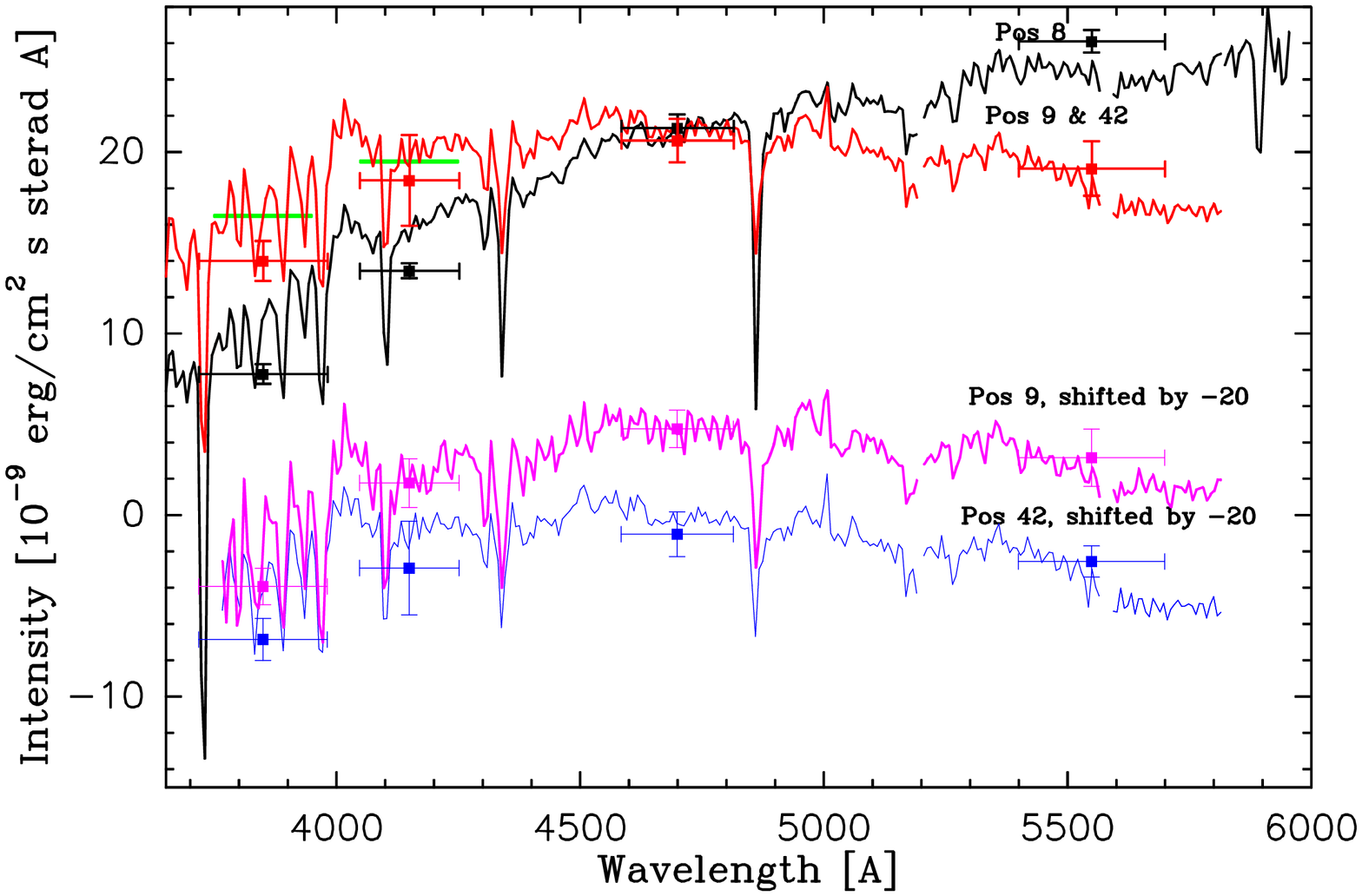}
\vspace{-20pt}
\caption{The observed surface brightness spectra for positions 8, 9 and 42, 
$\Delta I^{0}{\rm (Pos8 - OFF)}$, $\Delta I^{0}{\rm (Pos9 - OFF)}$, and $\Delta I^{0}{\rm (Pos42 - OFF)}$.
The black curve shows the overall mean for Pos8, including both the 'Master A\&B mean' and 'Secondary mean' 
spectra with weights of 0.4 and 0.6, respectively. The spectra for Pos9 and Pos42 are 
shown as magenta and blue curves, with their zero point shifted by -20 units.
The mean of the Pos9 and 42 spectra with weights 1 and 2.5, respectively, is shown as the red curve 
and is labelled as 'Pos 9 \& 42'. { The average fluxes of the 'Pos 9 \& 42' spectrum between 375 and 395 nm
and between 405 and 425 nm are indicated with green bars; they are used for the determination of the 400 nm
break strength.} The spectra have been smoothed with
a 5-pixel boxcar function and have a bin size of $\sim$0.7~nm and a resolution of $\sim$1.1~nm.
The spectra have gaps at the strong airglow lines at 520.0 and 557.7~nm. The intermediate band
photometric measurements (see Appendix C) are shown with the same colour coding as the spectra.
The filter half-widths and the photometric (statistical) standard errors are indicated with horizontal and
vertical bars. For the purpose of demonstrating the presence of the [O{\sc ii}]~372.8~nm line, 
in order to avoid confusion, the 365--380~nm section is shown only for the upper two spectra.
The spectra have been corrected for extinction and other atmospheric effects, i.e. they
refer to outside-the-atmosphere.}
\end{figure*}

\section{Resulting spectra}

The observational result consists of the surface brightness spectra at the positions 8, 9, and 42: 
$\Delta I^{0}$(Pos8 -- OFF), $\Delta I^{0}$(Pos9 -- OFF), and $\Delta I^{0}$(Pos42 -- OFF).
These spectra represent the surface brightness difference relative to 
the OFF-positions 18, 20, 24, 25, 32, 34, and 36; see Tables 1 and 2. The results refer to
the extinction-corrected, i.e. outside-the-atmosphere spectra.
The spectra for the IN-cloud positions, Pos9 and Pos42, have been calculated as:\\ 
$\Delta I^{0}$(Pos9 - OFF) = $\Delta I^{0}$(Pos8 - OFF) - $\Delta I^{0}$(Pos8 - Pos9), \\ 
$\Delta I^{0}$(Pos42 - OFF) = $\Delta I^{0}$(Pos8 -- OFF) - $\Delta I^{0}$(Pos8 - Pos42),
where $\Delta I^{0}$(Pos8 - Pos9) and $\Delta I^{0}$(Pos8 - Pos42) are the 
differential spectra w.r.t. Pos8 as obtained from observations and corrected
for extinction only (see Section 4.1). For  $\Delta I^{0}$(Pos8 - OFF) the mean of the 
'Master mean' and ``Secondary mean'' spectra was used in this case.

The results for Pos8 are shown in Fig. 5 
separately for the 'Master mean' and 'Secondary mean' spectra, as detailed in Table 2. 
Two different reduction methods for the ADL, methods A and B, as described in Appendix A.2.1 
and A.2.2, respectively, have been applied to the 'Master spectra'. 
The resulting mean spectra are shown in Fig. 5 as the lower pair of black (method A) and red (method B) 
curves, 'shifted by -10 units'. Their difference is shown as the lowest blue curve with the zero 
level indicated as red line. The results for methods A and B are seen to agree very closely 
which makes us confident that the effects of the AGL 
temporal and ADL spatial variations have been satisfactorily accounted for. 

For the red end of the spectra, at 582--595~nm, the results from only one 
night are included (Pos20, night 2010-12-14). 
Even in this quiescent night the variations of the atmospheric NaD line 
 and the adjacent OH(8-2) bands cause in this wavelength slot a substantially larger disturbance 
than is elsewhere the case. However, the detection of the NaD absorption line in 
the $\Delta I^{0}$(Pos8 - OFF) spectrum is well secured.

The mean 'Secondary spectrum', as described in Appendix A.2.3, is shown as the 
uppermost red curve in Fig. 5  together with the mean of the two (methods A and B) 
'Master mean' spectra (black curve). 
The difference between the two spectra is shown as the upper
blue curve, with its zero level indicated as the red line. 

An overall average of all Pos8 spectra, i.e. the mean of  the 'Master A\&B' and the 'Secondary 
mean' spectra is shown in Fig. 6 as the black curve. Before averaging, the 'Secondary mean' 
spectrum was adjusted to the 'Master mean' overall continuum by means of a 4$^{th}$ degree
polynomial fit. Thus, this spectrum has the same broad-band continuum as the 'Master mean' 
spectrum but an improved S/N ratio, corresponding to the combined integration time of  
the 'Master' and 'Secondary mean' spectra.

The spectra for Pos9 and Pos42 are 
shown in Fig. 6 as the~magenta and blue curves, with their zero point shifted by -20 units.
The weighted mean of Pos9 and Pos42 spectra (weights 1 and 2.5) is shown as the red line 
labelled with 'Pos 9 \& 42'.

An inspection of the spectra reveals the following salient features which
reflect the fact that the spectra are dominated by the scattered light from the
all-sky ISRF plus the line emission from ionized gas in the cloud area: \\
(i) the spectra show 
strong Fraunhofer lines: Ca{\sc ii}~H (396.8~nm, blended with H\,$\epsilon$), Ca{\sc ii}~K (323.3~nm), 
G~band (430.0~nm), Mg{\sc i}+MgH (518.0~nm), 
the Balmer H\,$\beta$ (486.1~nm), H\,$\gamma$ (434.1~nm), H\,$\delta$ (410.3~nm);\\
{ (ii) the 400~nm break, present in the integrated spectra of galaxies,
is also clearly visible in our scattered light spectra. The break strength, $D(4000)$, is defined 
as the ratio of the average flux density between 405 and 425~nm to that between 375 and 395~nm 
\citep{bruzual}, where the flux density is in units of erg s$^{-1}$cm$^{-2}$Hz$^{-1}$. For the 
'Pos 9\&42' spectrum the mean flux densities in the two reference intervals are 19.5 and 16.5 \cgs, and are 
indicated with green bars in Fig.~6. (In determining these mean fluxes we have taken into account
the small contribution by the LOS Balmer line emission of ionised gas, see Paper II, Section 2.1.3.)  
Transforming the flux densities to erg s$^{-1}$cm$^{-2}$Hz$^{-1}$ units their ratio is $D(4000) \approx 1.4$. 
This is at the lower end of the range of $D(4000)$ values found for the {\em total light} of spiral 
galaxies (see e.g \citealt{bruzual,hamilton,dressler}). This can be understood
because our spectra represent the light from the outer parts of the Milky Way Galaxy with little
contribution from the bulge component with its stronger 400 nm break. Because of the intermediate extinction
towards Pos9 and Pos42 the overall spectral shape is closely similar to integrated starlight spectrum. For the
strongly reddened Pos8 spectrum a similar $D(4000)$ strength is indicated, after the effect of reddening is 
accounted for.} \\
(iii) the SED of Pos 8 is substantially redder as compared to the SEDs of Pos9, 
Pos42 and the all-sky ISL (see Fig. 2). Because of the large optical depth for Pos8 at all 
wavelengths the SED shape is dominated by effects of multiple scattering and absorption;\\
(iv) the lower optical depth for Pos42 ($A_{\rm V} = 0.86$~mag)  as compared to 
Pos9 ($A_{\rm V} = 1.17$~mag) results in a lower 
surface brightness in visual and red, roughly proportional to their extinction ratio. 
At the blue end, because of saturation, the ratio is closer to 1;\\
(v) the Balmer lines, H\,$\beta$, H\,$\gamma$, and H\,$\delta$ are, especially
for Pos8, stronger than expected for scattered starlight (compare also with Fig. 2). 
This indicates that there is a substantial excess of line emission by ionized 
gas from the OFF positions. This is also the cause for the strong [O{\sc ii}] 372.8~nm line, seen
as an apparent absorption line in these spectra.

{ The spectra as displayed in Fig.~6 are provided also as machine readable files, see Appendix D
for explanations.} 
   
\begin{table*}
%\begin{centering}
\begin{minipage}{175mm}
\caption{Mean values and error estimates for selected wavelength intervals 
of the Pos8 and Pos42 spectra as specified in the text. The quantities given in the Table are:
mean value = average surface brightness over the wavelength interval;
$\sigma$(mean) = standard error of the mean value;
$\sigma$(pix-to-pix) = pixel-to-pixel standard error for
spectrum smoothed with 5-pixel boxcar function  resulting in $\sim 0.7$~nm bins.
 The unit is \cgs. }

\begin{tabular}{lllllllllllll}
\hline
   &      &  \multicolumn{11}{c}{Wavelength interval [nm]} \\ 
   &      &  \multicolumn{11}{c}{--------------------------------------------------------------------------------------------------} \\ 
   &      &380- & 385- & 391- & 401- & 416- & 438- & 470- & 494- & 510- & 534- & 565-\\
         &   &385  & 391  & 396  & 406  & 426  & 448  & 480  & 502  & 517  & 549  & 580 \\
\hline
   \multicolumn{13}{l}{$\Delta I^{0}$(Pos8 - OFF)}\\
'Master mean', method A &  mean     &9.5 & 9.8 & 12.3 & 16.0 & 16.8 & 18.3 & 21.4 & 22.8 & 22.2 & 24.8 & 24.5\\
 & $\sigma$(mean)&0.54 & 0.99 & 0.18 & 0.16 & 0.16 & 0.37 & 0.48 & 0.33 & 0.43 & 0.55 & 1.27\\
\hline
   \multicolumn{13}{l}{$\Delta I^{0}$(Pos8 - OFF)}\\ 
 'Master mean', method B & mean  & 9.7 & 10.1 & 12.5 & 16.5 & 17.2 & 18.6 & 21.6 & 23.0 & 22.0& 24.7& 23.8\\
  & $\sigma$(mean)& 1.15 & 1.22 & 0.57 & 0.83 & 0.56 & 0.73 & 0.69 & 0.65 & 0.58 & 0.39& 0.46\\
%  & $\sigma$(pix-to-pix)$^a$ &0.84 & 0.74 & 0.50 &0.29 & 0.25 & 0.21& 0.24 & 0.38& 0.25&0.11&0.18\\
\hline
   \multicolumn{13}{l}{$\Delta I^{0}$(Pos8 - OFF), 'Master Mean'}\\
 $\Delta I^{0}$(method A)- $\Delta I^{0}$(method B) & $\Delta$(mean) & -0.2 & -0.3 & -0.2 & -0.5  & -0.4 & -0.3 & -0.2 & -0.2 & 0.2 & 0.1 & 0.7 \\
\hline
   \multicolumn{13}{l}{$\Delta I^{0}$(Pos8 - OFF)}\\
%  &\multicolumn{12}{l}{$\Delta I^{0}$(Pos8-OFF), 'Master mean' {\em minus} 'Secondary mean' spectrum}\\
 $\Delta I^{0}$('Master') - $\Delta I^{0}$('Secondary') & $\sigma$(pix-to-pix)     
    &1.32&  1.15 & 0.79 &0.45 & 0.40 & 0.33& 0.38& 0.60& 0.41&0.18&0.28\\

$\frac{1}{2}[\Delta I^{0}$('Master')+$\Delta I^{0}$('Secondary')] & $\sigma$(pix-to-pix)\footnote{values are 
$\frac{1}{2}\times\sigma$(pix-to-pix) for $\Delta I^{0}$('Master')-$\Delta I^{0}$('Secondary')}
& 0.66& 0.58 & 0.40 &0.22  & 0.20 & 0.16& 0.19& 0.30& 0.20&0.09&0.14\\
\hline
 \multicolumn{13}{l} { $\Delta I^{0}$(Pos42-Pos8)]}\\
  & $\sigma$(mean)          & 0.51 &0.46  &0.46  &0.47  &0.47  &0.66  &0.81  &0.79  &0.66  &0.59  &1.15   \\
  & $\sigma$(pix-to-pix)    & 0.54 &0.41  &0.51  &0.26  &0.42  &0.38  &0.34  &0.51  &0.35  &0.48  &0.40  \\
\hline
\end{tabular}
\end{minipage}
\end{table*}

\begin{table*}
%\begin{centering}
\begin{minipage}{176mm}
\caption { Summary of the sky background brightness components towards our target area: 
the Zodiacal Light (ZL), atmospheric diffuse light (ADL) that is the sum of Airglow (AGL) and 
tropospheric scattered light, the scattered light from the cloud, $\Delta I^0$(Cloud  -- OFF), 
for Pos8, Pos9 or Pos42, the the Diffuse Galactic 
Light (DGL) in the surroundings (OFF positions), and our measurement of the EBL.
These values refer to the spectroscopic observations as listed in Table 2. 
Under 'differential corrections' the ranges of the corrections $\Delta I_{\rm ZL}$,
$\Delta I_{\rm ADL}$ and $\Delta I_{\rm stray}$ are given  for the different Pos8 -- OFF spectra 
listed in Table 2. 
A compilation of error estimates according to Table 3 and Sections
8.1 and 8.2 is included in the bottom part of the table. The calibration errors are in per cent.
All values refer to outside the atmosphere, except for $I_{\rm ADL}$ which is the value below the 
troposphere, at ground-level.  }
\begin{centering}
\begin{tabular}{lccl}

\hline
Component & Wavelength & Value & Reference/Comment  \\
          &  [nm]      &[\cgs\,]&                  \\
%          & $[$nm$]$   &  $[$ \cgs $]$ &            \\
\hline 
$I_{\rm ZL}$            & 500     & 102 -- 105 & \citet{Leinert} \\
$I_{\rm ZL}$ \footnote{derived from $I_{\rm ZL}$ at 500 nm using the Solar spectrum of \citet{Kur} 
and ZL colour of \citet{Leinert}, see Appendix A1}&400 -- 420 & 88 -- 99 & derived from $I_{\rm ZL}$ at 500 nm \\
$I_{\rm ADL}$           & 400 -- 420 & 83  -- 138 & see Section 7, equation (5) \\
$\Delta I^0$(Cloud -- OFF) \footnote{scattered light excess from the cloud, designated
$\Delta I^{0}$(Pos8 -- OFF),  $\Delta I^{0}$(Pos9 -- OFF), $\Delta I^{0}$(Pos42 -- OFF) in Section 8} 
&400 -- 420& 16 -- 22  & see Table 3  \\
$I_{\rm DGL}$(OFF)\footnote{scattered light in OFF areas} & 555 & 3.3 & see Paper II, Section 2.1.2  \\
$I_{\rm DGL}$(OFF)\footnote{derived from $I_{\rm DGL}$(OFF) at 555 nm} & 400 -- 420 &4.6& see Paper II, equation (7) \\
$I_{\rm EBL}$           & 400        &2.9& see Paper II \\
\multicolumn{4}{l}{\em Differential corrections}\\
$\Delta I_{\rm ZL}$     & 400 -- 420 &-2.7 -- 3.5 & see Table 2, Column (8)\\
$\Delta I_{\rm ADL}$    & 400 -- 420 & -1.8 -- 1.8 & see Appendix A2 and equation (A3)\\
$\Delta I_{\rm stray}$  &400 -- 500  & -0.6 -- 1.7 & Appendix B, Table B1\\
\hline
\multicolumn{4}{l}{\em Statistical errors}\\
$\sigma$(pixel-to-pixel)   &385 -- 400  & $\le 0.6$  & Section 8.1 and Table 3\\
$\sigma$(pixel-to-pixel)   &400 -- 500  & $\le 0.2$  & Section 8.1 and Table 3 \\
$\sigma$(mean)             &350 -- 500  & $\le 0.5$  & Section 8.1 and Table 3\\
\multicolumn{4}{l}{\em Calibration errors}\\
$\sigma$(stars)&           & $\le$ 7\% & Section 8.2   \\
$\sigma$(extinction)&           & $\le$ 1.2\% &  Section 8.2 \\
$\sigma$(aperture correction))&    & $\le$ 6\% & Section 8.2  \\
\hline
\end{tabular}
\end{centering}
\end{minipage}
%\end{centering}
\end{table*}

\subsection{Statistical errors}

The statistical errors influencing the observed spectra are considered
from two points of view:\\
(i) The pixel-to-pixel statistical error. It is mainly caused by the photon statistics
and instrumental read-out noise; this noise is {an} important limiting factor for the determination of
 the Fraunhofer line depths and the 400~nm discontinuity, which are utilized for the evaluation
of the scattered starlight contribution to the spectra.\\
(ii) The error in the zero level. Besides by photon statistics and read-out noise as in (i),
it is caused also by the uncertainties of the differential ADL and ZL corrections.

For the 'Master mean' spectrum the deviations of the four individual spectra from the mean
have been used to estimate the statistical error of the zero level. In the first two parts of Table 3 
the mean values and their mean errors, $\sigma$(mean), are given for selected wavelength intervals. 
The wavelength intervals are broad enough so that the $\sigma$(mean) values are dominated 
by the zero-level error and not %significantly influenced by 
the pixel-to-pixel noise.
The mean values for methods A and B  agree to within $\la$0.5\,\cgs\, 
(see third part of Table 3 and Fig. 5). This is consistent with the  $\sigma$(mean) 
values that for method A are  $\la$0.5\,\cgs\, over most of the wavelength range and it 
indicates that the differential ADL corrections do not introduce zero-level errors 
larger than this.
The smaller $\sigma$(mean) values for method A as compared to method B indicate that the AGL 
time variations (whose influence is minimized in method A) are more important as 
 an error source than the ADL zenith distance dependence (eliminated in method B). 

The pixel-to-pixel noise cannot be directly  determined from the mean 
spectra because of the intrinsic structures in the underlying scattered
light (ISL) spectrum. We have therefore used the difference
spectrum 'Master mean (A\&B)' { minus} 'Secondary mean',  
$\Delta I^{0}{\rm('Master')} - \Delta I^{0}{\rm('Secondary')}$  (see  Fig. 5)
to derive the   $\sigma$(pixel-to-pixel) values as listed in the fourth part of Table 3.
Because the total integration times of the 'Master' and 'Secondary mean' 
spectra are similar, the noise for each of them is $\sim 1/\sqrt 2$ times, and the  noise 
for their mean is  $\sim 1/2$ times the $\sigma$(pixel-to-pixel) 
noise of the difference spectrum.
The  $\sigma$(pixel-to-pixel) values  for the mean are also given in part 4 of of Table 3.

The values of $\sigma$(mean) for  the   $\Delta I^{0}$(Pos42 - Pos8) 
spectrum (Table 3, Part 5) were derived from the deviations of the individual spectra from their mean.
Since the contribution by the underlying  spectral structure of the ISL spectrum
turned out to be a minor factor compared to the other noise components, 
the  $\sigma$(pixel-to-pixel) values of the $\Delta I^{0}$(Pos42 - Pos8)   
 spectrum were directly adopted. For the error estimates of the mean 
of $\Delta I^{0}$(Pos9 - Pos8) and $\Delta I^{0}$(Pos42 - Pos8) we have adopted the 
values as derived for Pos42 because it dominates the mean by its larger weight (2.5 vs. 1).

 In summary, the pixel-to-pixel statistical errors, important for the fitting accuracy 
(and elimination) of the scattered light spectrum, are $\la 0.6$\,\cgs\, for  
$\lambda < 400$nm, and  $\la 0.2$\,\cgs\, for $\lambda > 400$~nm,
respectively, as found for the mean of the 'Master' and 'Secondary' mean spectra.
The accuracy of the zero level of the 'Master Mean' spectrum can conservatively be estimated
to be $\leq 0.5$\,\cgs\, for $\lambda \geq 390$nm as indicated by the $\sigma$(mean)$\leq 0.5$\,\cgs\,  
errors for the 'Master mean' method A spectrum, and by the  $\leq 0.4$\,\cgs\, differences between the
'Master mean' method A and B spectra.

\subsection{Calibration errors}

Our observations are differential and the night sky component separation   
is carried out using data from the same telescope for all components.
Therefore, unlike in the methods applied by \citet{b1} and \citet{Matsu11},  
an absolute flux calibration of high accuracy is not required for the component separation. 
The calibration error appears essentially only as a scaling factor in the derived EBL value.
  
A source of calibration errors are the observations of standard stars, 
i.e. their pure observational errors, the photometric stability of the instrumental system over up to
two days, and the transformation to the standard spectrophotometric system. 
For the service mode observations the VLT/FORS User 
Manual\footnote{http://www.eso.org/sci/facilities/paranal/instruments/\\\~/fors/doc.html; 
VLT-MAN-ESO-13100-1543\_v82.pdf; Table 4.1} cites an accuracy 
of 10\% for an individual night. For our Master mean' and 'Secondary mean' spectra,
which combine averages of three or six nights, respectively, we somewhat arbitrarily estimate 
that this error is reduced to $\la$7\%.
 
The error caused by uncertainties of extinction correction has been estimated to be 
$< 1.2$\% (see Section 6.3).  

In addition, there are the errors caused by the specific
aspects of extended surface brightness calibration. These are the errors for the
spectrometer solid angle $\Omega$ of $\pm1$\%,  and of the aperture correction factor $T(A)$
of  $\pm6$\%, as discussed in Sections 6.4 and 6.5.

A quadratic addition of the standard star photometric, the extinction, and the aperture correction 
factor errors gives the overall calibration error estimate of  $\la \pm9$\%.

For comparison, we show in Fig. 6 also results of the intermediate band photometry (see Appendix C).
The agreement between the spectroscopic and photometric values is within the  $\la \pm9$\% calibration
uncertainty. 

\subsection{Summary of sky components, differential corrections and errors}

{ We summarise in Table~4 the contributions of the sky surface brightness components 
valid for the sky area and time slots of our spectroscopic observations: the Zodiacal Light, 
Atmospheric Diffuse Light and Diffuse Galactic Light. For comparison, the surface brightness
excess of the positions 8, 9 and 42 in L~1642 and our result for EBL intensity are given.
The range of values of the small differential corrections caused by the ZL, ADL and instrumental
straylight are listed next; we refer to the detailed description of these corrections in
Section 7, Appendix A and B. In the mean spectrum, $\Delta I^{0}$(Pos8 -- OFF), the corrections 
with opposite signs cancel out partly. 

In the lower part of the table we summarise the errors; the detailed
description of them is given in Sections 8.1 and 8.2.

The general DGL, i.e. the scattered light from dust in the surroungs of the cloud, has been
estimated using the mean extinction for the OFF positions , $A_{\rm V}\approx 0.15$~mag (see Section~3
and Table~1). Using the relationship between the optical extinction and 200~$\mu$m
surface brightness on the one, and that  between the optical and 200~$\mu$m surface 
brightnesses on the other side (see Appendix C) this $A_{\rm V}$ value is found to correspond
to a DGL value of $I_{\rm DGL} = 3.3$ \cgs\, at $\lambda = 555$ nm. For details, see Paper II,
Section 2.1.2. }

\section{Summary and Conclusions}
We utilize the shadowing effect of an opaque dark cloud to measure the EBL plus possible 
other diffuse light components originating from beyond the cloud. However, besides screening background 
light the dark cloud also scatters light from the all-sky Interstellar Radiation Field (ISRF, consisting mainly 
of ISL) into the line of sight. In order to separate the EBL and ISL components we utilize the fact 
that their spectra  are different: smooth continuum (EBL) vs. line spectrum (ISL). The component 
separation and results for the EBL are presented in the accompanying Paper~II \citep{mat17b}. 
   
In this paper we have presented spectrophotometric measurements of the faint surface brightness in the area
of the high galactic latitude dark cloud Lynds 1642. The cloud has a high opacity core with 
$A_{\rm V}\ga 15$~mag, blocking the EBL almost completely, while there are 'clear' areas within 2\fdg5 
from the cloud core with good transparency ($A_V\approx 0.15$~mag). We have presented 
a spectrum for the difference 'opaque core minus clear sky'. 
In addition, two intermediate opacity positions, $A_{\rm V} \approx 1$~mag, have been observed 
and their spectra 'cloud minus clear sky' will be utilized in Paper~II to 
facilitate the separation of the ISL and EBL light components. Because of 
the pseudosimultaneous differential measurements the spectra are virtually free of the large foreground
atmospheric and Zodiacal Light components.

For the separation of EBL and scattered ISL a spectral 
resolution of $R\sim 400$ is required and was achieved with long slit spectroscopy with VLT/FORS. 
A precision of $\sim$\,0.5\,\cgs\, was achieved for the averaged differential spectrum 'cloud core minus
clear sky'. Such a precision is required for the study of the EBL signal which is in the range of  
1 to a few times \cgs\,.
In the dark cloud method all diffuse light components are measured using the same equipment, 
and the component separation is done before the calibration. Thus, we do not need an unusually 
high {\em absolute} calibration accuracy as is the case for projects in which 
the small EBL signal has to be derived as the difference of $\sim$50 to 100 times larger surface 
brightness signals. 

An inspection of the spectra (see Fig. 6) reveals the following salient features which
reflect the fact that the spectra are dominated by scattered light from the
all-sky ISRF plus line emission from ionized gas in the cloud area: 
(i) the spectra show distinctly the strong Fraunhofer lines: Ca{\sc ii}~H (blended with  H$\epsilon$), 
Ca{\sc ii}~K, G~band, Mg{\sc i}+MgH, the Balmer H\,$\beta$, H\,$\gamma$, H\,$\delta$, and the 400~nm 
discontinuity, typical of integrated spectra of late type spiral galaxies;
(ii) the SED of the opaque core is substantially reddened as compared to the SEDs of the
intermediate opacity positions and the ISL. Because of the large optical depth of the core 
the SED shape is dominated by effects of multiple scattering and absorption;
(iii) the Balmer lines are stronger than expected for scattered starlight. 
This indicates that there is a substantial excess of line emission by ionized 
gas from the OFF positions. This is also the cause for the strong [O{\sc ii}] 372.8~nm line, seen
as an apparent absorption line in these spectra. On the other hand, [O{\sc iii}] 500.7~nm  is seen as an 
emission line indicating that it originates mainly as scattered light from the all-sky ISRF.

{ The separation of the EBL from the scattered light of the dark nebula is
described in the accompanying Paper~II \citep{mat17b}. 
 As template for the scattered starlight 
spectrum we make use of the spectra at the two semi-transparent positions 9 and 42. 
The main results are: The  EBL intensity at 400~nm  is
$I_{\rm EBL}=2.9\pm1.1$\,\cgs\, or $11.6\pm4.4$\,nW m$^{-2}$sr$^{-1}$,  
which represents a 2.6$\sigma$ detection; the scaling uncertainty is +20\%/-16\%. 
At 520~nm we have set a 2$\sigma$ upper limit of $I_{\rm EBL} \le$4.5\,\cgs\, 
or  $\le $ 23.4~nW m$^{-2}$sr$^{-1}$ +20\%/-16\% .
Our EBL value at 400~nm is $\ga 2$ times as high as the integrated light of galaxies.
No known diffuse light sources, such as light from Milky Way halo, intra-cluster or intra-group stars
appear capable of explaining the observed EBL excess over the integrated light of galaxies.

Besides the main purpose, i.e. the EBL determination, the resulting spectra of the present paper 
can also be used for two other purposes:}
(i) determination of the scattering properties of dust, namely the albedo and the 
asymmetry parameter as function of wavelength (see e.g. \citealt{mat70, lau}), and (ii) empirical
test of the synthetic spectrum of the Solar neighbourhood ISL (see Paper~II, Appendix A). 
Concerning the latter application we note especially that the dark cloud 'sees' the Milky Way from
a vantage point that is $\sim$\,85~pc off the galactic plane.

%\newpage

\section*{Acknowledgements}
We thank the ESO staff at Paranal and in Garching for their excellent service 
{ and an anonymous referee for suggesting several improvements to the paper.}
This research has made use of the USNOFS Image and Catalogue~archive
   operated by the United States Naval Observatory, Flagstaff Station
   (http://www.nofs.navy.mil/data/fchpix/).\\
The Digitized Sky Surveys were produced at the Space Telescope Science Institute 
under U.S. Government grant NAG W-2166. The images of these surveys are based on 
photographic data obtained using the Oschin Schmidt Telescope on Palomar Mountain 
and the UK Schmidt Telescope. The plates were processed into the present compressed 
digital form with the permission of these institutions. KM and KL acknowledge the 
support from the Research Council for Natural Sciences and Engineering (Finland) and PV  
acknowledges the support from the National Research Foundation of South Africa.

\newpage

\appendix

\section{Differential corrections for Zodiacal Light and Atmospheric Diffuse light}

\subsection{Differential Zodiacal Light correction}

With the designation
$ \Delta I_{\rm ZL}=  I_{\rm ZL}({\rm Pos8})- I_{\rm ZL}({\rm PosN})$ 
the ZL correction term in equation (7), reduced to outside the atmosphere, can be expressed as:
\begin{eqnarray}
\lefteqn{\Delta I^{\rm {\rm corr}}_{\rm ZL} = } \nonumber\\
\lefteqn{=[I_{\rm ZL}{\rm (Pos8)} e^{-\tau(\lambda) (X+\Delta X)} - I_{\rm ZL}{\rm (PosN)} e^{-\tau(\lambda) X}]e^{\tau(\lambda) X} } \nonumber\\ 
\lefteqn{= \Delta I_{\rm ZL} -\Delta X \tau(\lambda) I_{\rm ZL}} 
\end{eqnarray}
where the term $\Delta X \tau(\lambda) I_{\rm ZL}$ accounts for the slightly different atmospheric 
extinctions for the Pos8 and PosN observations.
With the ecliptic coordinates $\lambda-\lambda_{\sun}$ and $\beta$ given for the observed 
ON and OFF positions at each observing date, we have interpolated 
from Table 17 of \citet{Leinert} the mean ZL intensity for Po8 and PosN, $I_{\rm ZL}$, 
and the difference $ \Delta I_{\rm ZL}$.
Their values at $\lambda = 500$~nm are given in column (8) of Table 2. 
While the absolute ZL intensities are known only to
an accuracy of $\sim 10$\,\cgs\, \citep{Leinert} the 
differences  $\Delta I_{\rm ZL}$ over separations of a few degrees are more accurate,
of the order of a few tenths of \cgs\,,  as estimated from Table 16 of \citet{Leinert}.

In order to estimate $I_{\rm ZL}(\lambda)$ and $ \Delta I_{\rm ZL}$ at other wavelengths 
than 500~nm we adopt  as starting point the Solar spectrum as 
given by \citet{Kur} (in digital form at {\tt http://kurucz.harvard.edu/sun.html}). 
To account for the redder colour of the ZL as 
compared to the Sun we multiply the Solar spectrum with the correction factor $f_{\rm co-90}$ as 
given by equation (22) in  \citet{Leinert}. For the wavelength range  $\lambda = 360-580$~nm it can
be satisfactorily represented by: $f_{\rm co-90}(\lambda)=1.0 + 0.74 \cdot 10^{-3}(\lambda-500$~nm).
The solar and the
ZL spectra are rather similar to the ISL spectrum. Therefore, the ZL correction is
partly 'absorbed' into the scattered ISL component and its uncertainty is not 
causing an error in the EBL value with full weight (see Paper~II).

\subsection{Differential Atmospheric Diffuse Light correction}

The AGL and atmospheric scattered light cause both random and systematic variations of 
the sky brightness. The random effects consist of\\
(i) temporary variations of the AGL on time scales of $\sim$10 minutes to hours, and \\
(ii) irregular ('cloudy')
spatial variations of the AGL on scales from several degrees upwards. \\
The systematic effect consists of \\
(iii) the zenith distance dependence of AGL and tropospheric scattered light, jointly 
called atmospheric diffuse light, ADL. %proportional to the airmass.

For the ADL correction of the 'Master spectra' we have used two independent complementary methods.\\

\subsubsection{Method A that minimizes the effect of AGL time variations}

The observations in the cycle Pos$8_1$ - PosN - Pos$8_2$  are equally spaced in time. 
The airglow time variations are thus eliminated as long as they are linear in time, and 
the error introduced by the non-linearity is expected to be substantially smaller than 
$|I_1(Pos8)-I_2(Pos8)|$. However, because the three 
observations, Pos$8_1$, PosN, and Pos$8_2$, are taken at slightly different zenith distances,
a systematic effect is caused by the zenith distance dependence of the ADL.

Near the zenith the atmosphere can be approximated as plane parallel, 
and at airmasses  $X \la 1.3$ the ADL intensity
can be given by
\begin{equation}
I^{\rm X}_{\rm ADL}= I^X_{\rm AGL}+I^X_{\rm SCA} = I^{1.0}_{\rm ADL}[1 + k(\lambda)(X-1)]
\end{equation}
where $I^{1.0}_{\rm ADL}$ is the ADL intensity in zenith ($X = 1.0$). 
The ADL difference between airmasses  $X + \Delta X$ and  $X$ is thus given by 
\begin{equation}
\Delta I^{X+\Delta X, X}_{\rm{ADL}}  = {\scriptstyle\frac{k(\lambda)}{1+k(\lambda)(X-1)}} \Delta X\cdot I^X_{\rm {ADL}} 
   =  C(\lambda)\cdot \Delta X\cdot I^X_{\rm ADL} 
\end{equation}
where $C(\lambda)= k(\lambda)[1+k(\lambda)(X-1)]^{-1}$.
We have determined the values of $k(\lambda)$ and  $C(\lambda)$ using the
tables of \citet{Ashburn} that are based on the solution of the multiple 
scattering problem in plane parallel atmosphere using the Chandrasekhar-formalism
\citep{chandra}. The Ashburn tables give the combined AGL and tropospheric scattered
light at ground for different extinctions and AGL layer heights. The case of infinite layer
height can be used for modelling the scattered light from extraterrestrial light sources.
Combining the results for the layer heights of 100 km (AGL) and $\infty$ (ZL and ISL)
we obtained the values  $k(\lambda)$ = 0.65, 0.70, 0.75, 0.80, 0.85 and  
$C(\lambda)$ = 0.61, 0.65, 0.70, 0.74, 0.78 at $\lambda$ = 360, 400, 450, 500 and 550-600~nm, 
respectively. The values of  $C(\lambda)$ are for an airmass of 1.10, representative 
for the 'Master spectra'.

Using  equation (4) we obtain  for the ADL correction, referred to outside the atmosphere, the expression
\begin{equation}
\Delta I_{\rm ADL}^{\rm corr}= [I^X_{\rm obs} -I_{\rm ZL}e^{-\tau(\lambda) X}]C(\lambda) \Delta X e^{\tau(\lambda) X}
\end{equation}

Applying the differential ADL and ZL corrections, the outside-the-atmosphere ON minus OFF surface 
brightness difference is then given according to equation (7)
\begin{eqnarray}
\Delta I^{0}{\rm (Pos8-PosN)} & = & \Delta I_{\rm obs}{\rm (Pos8-PosN)}e^{\tau(\lambda) X} -  \nonumber \\ 
                                 & - &  \Delta I_{\rm ADL}^{\rm corr}-\Delta I_{\rm ZL}^{\rm corr}
\end{eqnarray}

We give in column (7) of Table 2 for each measurement the airmass for the OFF position (PosN) and 
the airmass difference $\Delta X = X$(Pos8) - $X$(PosN).
For airmass differences of $\Delta X =$ -0.034 to +0.032 the corrections 
$\Delta I^{X,X+\Delta X}_{\rm ADL}$ range from  -1.8 to +1.8\,\cgs.
The correction for the 'Master mean' spectrum is +0.8\,\cgs.\\

\subsubsection{Method B that avoids the effect of ADL zenith distance dependence}

We have used also another reduction method in which
the ADL zenith distance dependence does not enter. In the measuring cycle Pos8$_1$ - PosN - Pos8$_2$ 
the zenith distances for Pos8$_1$ and Pos8$_2$ bracket that of the PosN measurement. 
It is possible to choose the weights $w_1$ and $w_2$ in such a way that for the weighted mean  
$\overline{I{\rm (Pos8)}} = w_1 I_1{\rm (Pos8)}+ w_2 I_2{\rm (Pos8)}$ the weighted airmass 
$\overline{X{\rm (Pos8)}} = w_1 X_1{\rm (Pos8)}+ w_2 X_2{\rm (Pos8)}$ is equal to the airmass of the 
PosN measurement. The weights vary between 0.46/0.54  
and 0.18/0.82. The observed surface brightness difference \\
$\Delta I_{\rm obs}{\rm (Pos8-PosN)}=\overline{I{\rm (Pos8)}}-I{\rm (PosN)}$
has to be corrected in this case for ZL only, and we have instead of equation (A5):
\begin{eqnarray}
\Delta I^{0}{\rm (Pos8-PosN)}  =  \Delta I_{\rm obs}{\rm (Pos8-PosN)}e^{\tau(\lambda) X} - \Delta I_{\rm ZL} \nonumber \\
\end{eqnarray}
A drawback of this method is that the elimination of the AGL time variations is 
less optimal.

We have applied both methods, A and B, 
to the correction of ``Master spectra''. The results are presented in Section 8 and Fig. 5.\\

\subsubsection{ADL correction for the 'Secondary spectra'}

For the 'secondary spectra' the AGL time variation was too large, 
 $|I_1{\rm (Pos8)}-I_2{\rm (Pos8)}| > 5$\,\cgs\,, to enable an independent accurate extraction
of the differential spectra $\Delta I_{\rm obs}$(Pos8 -- PosN). Using Method A
as described above the resulting spectrum still contained a substantial residual
component of the AGL which had not been removed by the linear interpolation.
We can assume, however, that the spectral form of this residual AGL component
is well approximated by the spectrum of the difference signal between the two consecutive
observations of the same position, $\Delta I_{\rm AGL}$(Pos8$_{1-2}$) = $I_1$(Pos8) -- $I_2$(Pos8).
Thus, we have added to the spectrum $\Delta I_{\rm obs}$(Pos8 -- PosN)
%, derived using method A, 
a suitable fraction, $-0.5<F<0.5$, of the differential AGL spectrum so that the resulting corrected spectrum 
\begin{eqnarray} 
\lefteqn{\Delta I^{0}{\rm (Pos8 - PosN)} ={} } \nonumber \\
                          &  &  {}= \Delta I_{\rm obs}{\rm (Pos8 - PosN)} + F\cdot \Delta I_{\rm AGL}{\rm (Pos8}_{1-2})
\end{eqnarray}
is optimised in two ways: (i) it fits as well as possible to the mean of 'Master spectra', and
(ii) the residual contributions of the AGL emission bands as seen in the difference spectrum 
$\Delta I_{\rm AGL}{\rm (Pos8}_{1-2})$ are minimised. We notice that this correction procedure does not
influence the strengths of the non-AGL related features in the  $\Delta I_{\rm obs}{\rm (Pos8 - PosN)}$ spectrum. 
Since the mean 'Secondary spectrum' is mainly used to improve the S/N ratio, we adjust, 
as a final step, its continuum level by a polynomial fit to the mean 'Master spectrum'.
For the result see Section 8 and Fig. 5.

\section{Instrumental straylight}

\subsection{Measurement of straylight profile of a star}

The stray radiation profile of a star, $I_{\rm stray}(r)$, in the range of $r\approx1$~arcmin -- 1 deg 
is thought to be caused mainly by scattering from the telescope main mirror 
micro-ripple and dust contamination (see e.g. \citealt{Beckers}).

Using the same LSS setup with FORS1 at UT2 as described in Section 4.2  
we have measured the straylight over 
360 - 580~nm. The measurements were made on 2009-01-26 using 
Sirius as light source.  The angular range of  $r$ = 100\arcsec - 1400\arcsec\, was 
covered with three slit positions, 45\arcsec\, East of Sirius, oriented along the N-S direction:   
$r$ =  $\sim$100\arcsec\, -- 500\arcsec\, South; $\sim$400\arcsec - 800\arcsec\, North; 
and  $\sim$1000\arcsec\, -- 1400\arcsec\, South. As zero level we used spectra taken 
symmetrically at 2\degr\, North and South. Integration time was 255 s and airmass 1.0.
The data reduction and calibration were performed as described in Sections 5 and 6.
The scaling of the middle region was adjusted slightly ($\sim12\%$) to make 
it fit smoothly to the innermost region in the overlapping part at 400\arcsec\, - 500\arcsec\, 
and, correspondingly, the outermost region was adjusted to the middle region (by $\la10\%$) 
to provide a smooth continuation over the gap from $\sim$800\arcsec\, to 1000\arcsec\,. 
The closely linear form of ~~ log$I_{\rm stray}(r)$ vs. log$r$ ~~ enabled 
a safe interpolation over 800\arcsec - 1000\arcsec\, as well as the
the small gaps between Chip\,1 and 2.
Spikes by stars were removed by interpolation of the intensity 
 over such locations. 
 Finally, the straylight spectra were divided by the
spectra of Sirius, adopted from \citet{Kur}, convolved to the resolution of our straylight 
spectra, and extinction corrected to below the atmosphere. 

We show in Fig. B1 the straylight intensity $I_{\rm stray}(r)$ as function of offset 
from Sirius, $r$, for four wavelength slots,  $\lambda=360-380$~nm, $400-420$~nm, $480-500$~nm  and 
$560-580$~nm. The observed straylight follows  over the range 
120\arcsec\, - 1400\arcsec\, the functional form according to \citet{King}, 
$I_{\rm stray} \propto r^{-2}$,  with a slight upturn at  $r<$120\arcsec\,.
For the wavelength bands $420-460$~nm and $480-520$~nm 
corresponding to the Tycho $B_T$ and $V_T$ magnitudes,
we obtain the straight line fits:\\
log\,$I_{\rm stray} = -1.988(\pm0.003) {\rm log} r\arcsec - 2.837(\pm0.008)$ and\\ 
log\,$I_{\rm stray} = -1.985(\pm0.004) {\rm log} r\arcsec - 2.879(\pm0.012)$.\\ 
Our $I_{\rm stray}$  level at 420--460~nm is by  $\sim$0.50~mag, i.e by a factor of $\sim${1.60} 
higher than the straylight profile of King, valid for the blue (photographic) band,
shown in Fig. B1 for reference.

\subsection{Straylight for observed positions}

The  straylight intensities were calculated in the  $B_T$ and $V_T$ bands 
for each target position in the L\,1642 cloud area. 
For stars with $B_T \le 12.0$~mag we adopted $B_T$ and $V_T$~magnitudes from
the Tycho-2 catalogue\footnote{http://archive.eso.org/skycat/servers/ASTROM}
\citep{Hog} up to an offset of 180~arcmin\,; for the fainter stars with $B > 12.0$~mag 
we adopted the $B$ and $V$ magnitudes from The Naval Observatory Merged Astrometric 
Dataset (NOMAD) catalogue\footnote{http://www.nofs.navy.mil/nomad/} 
(\citealt{Nomad04, Nomad05}) up to an offset of 100~arcmin\,.
The straylight corrections are given in Table B1, separately 
for Tycho-2 and Nomad stars and for the two wavelength bands $B$ and $V$. 

\begin{table*}
\centering
\caption{Instrumental straylight intensities, $I_{\rm stray}$, in the $B$ and $V$ bands 
from stars in the L\,1642 cloud area. The unit is S$_{10}$ (one $10^{th}$~magnitude star per $\sq\degr$).
{1 S$_{10}$ corresponds to 2.17\,\cgs\, in the $B$ band and to 1.18\,\cgs\, in the $V$ band.}
Columns 2 and 3 give the straylight for the bright stars, $B_T \le 12$~mag, using Tycho-2 magnitudes, and
columns 4 and 5 for the fainter stars using magnitudes from the Nomad catalogue. Columns 6 and 7 give the 
(differential) straylight caused by the enhanced Diffuse Galactic Light in the L\,1642 cloud centre, and columns
8 and 9 the sum of these straylight components for Pos8 and the 'Master' and 'Secondary' OFF position mean values .}
\begin{tabular}{lccccrrrrcc}
\hline
Position & \multicolumn{8}{c}{$I_{\rm stray}$ [S$_{10}$]} & \multicolumn{2}{c}{Weights}\\
         & \multicolumn{8}{c}{---------------------------------------------------------------------------------------------}
         & \multicolumn{2}{c}{for mean spectra} \\
         & \multicolumn{2}{c}{Tycho-2 $B_T \le 12^m$} &\multicolumn{2}{c}{Nomad $B > 12^m$}
         &\multicolumn{2}{c}{DGL excess}
         & \multicolumn{2}{c}{Total straylight}& & \\
         & $B_T$ &  $V_T$ & B~~ &  V~~ &B~~ & V~~ & B~~ & V~~ & Master & Secondary\\
\hline
POS18 & 1.17 & 2.33 & 0.52 & 0.82&&&&&0.5& 0.7\\
POS20 & 0.66 & 1.18 & 0.49 & 0.66&&&&&0.5& - \\
POS24 & 0.33 & 0.72 & 0.31 & 0.51&&&&&2.5& 1 \\
POS25 & 0.36 & 0.74 & 0.38 & 0.61&&&&&-  &0.5\\
POS32 & 0.44 & 0.93 & 0.37 & 0.66&&&&&-  &0.5\\
POS34 & 0.54 & 1.30 & 0.46 & 0.71&&&&&-  &2 \\
POS36 & 0.42 & 0.92 & 0.40 & 0.61&&&&&-  & 1\\
   &       &       &       &          &  \\
Master mean    &0.49 & 1.02 & 0.36 & 0.58 &$<0.01$ &$<0.01$&0.85&1.60&3.5 &  \\
Secondary mean &0.44 & 1.02 & 0.40 & 0.64 &$<0.01$&$<0.01$&0.84&1.66&     &5.7 \\ 
Master mean - Pos8 &-0.08 & -0.28 &0.12 & 0.21 &-0.10 &-0.14&-0.06&-0.21&3.5 &  \\
Secondary mean - Pos 8&-0.12 &-0.28 & 0.16 & 0.27 &-0.10&-0.14&-0.07&-0.15&     &5.7 \\ 
\hline
POS08  & 0.57 & 1.30 & 0.24 & 0.37&0.10&0.14&0.91&1.81&-&-\\   
POS09  & 0.55 & 1.19 & 0.28 & 0.46&&&&&-&-\\
POS42 & 0.51 & 1.15 & 0.33 & 0.49&&&&&-&-\\
\hline
\vspace{0.5cm}
\end{tabular}
\end{table*}

Besides the stars also the excess scattered light brightness in the L\,1642 cloud area
(that can be considered as DGL enhancement)
causes a small contribution to the differential straylight correction. While 
the straylight from fainter stars ($B > 12$~mag) is reduced for positions in the 
obscured cloud area, this effect is partly compensated by the additional straylight
originating from the enhanced surface brightness.
The distribution of the optical surface brightness in L\,1642 \citep{lau} 
follows closely the 100~$\mu$m and 200 $\mu$m distributions \citep{Lehtinen04,Lehtinen07}. 
Thus, we used the  ISOPHOT 200~$\mu$m map, covering an area 
of $\sim$1.2~$\sq\degr$ as a proxy for the optical $B$ and $V$ band surface brightness 
distributions. Using our optical and ISOPHOT  200~$\mu$m  photometry
we have derived a relationship for the conversion from the 200 $\mu$m  
to the $B$ and $V$ band surface brightness (see Appendix C.1).
The resulting straylight for Pos8 was found to be 0.10 and 0.14 S$_{10}$
or  0.22 and 0.17\,\cgs\,, in $B$ and $V$, respectively, 
while the values for the OFF cloud positions were $<0.01$ S$_{10}$.

The spectrum to be used for the EBL determination is the mean of the individual
differential spectra.  
Therefore, instead of the straylight corrections for individual positions,  
the mean value of the differential corrections 
is what matters. Because of the different weights of the individual  spectra in the 
mean spectrum also the corrections have to be weighted correspondingly. In Table B1 we give
the weighted mean straylight corrections for the 'Master' and the 'Secondary' 
mean spectrum. 

The corrections for the 'Master Spectrum' in the $B$ band are -0.08, 0.12, and 
-0.10~S$_{10}$ for Tycho-2 stars, Nomad stars and the excess DGL brightness, respectively, 
resulting in a total of -0.06~S~$_{10}$ or -0.13\,\cgs\,.

In the $V$ band the corresponding corrections are 
-0.28, 0.21, and -0.14~S$_{10}$, in total -0.21~S$_{10}$ or -0.25~\cgs\,. 
Summing up the corrections for the 'Secondary mean' spectrum we obtain -0.07~S$_{10}$ 
in $B$ and -0.15~S$_{10}$ in $V$, or -0.15 and $-0.18$~\cgs\,, respectively.

In conclusion, the straylight corrections are small enough to be neglected in the 
further analysis. No differential straylight corrections are needed for Pos9 and Pos42, either.
Also, any uncertainties due to changes of the straylight level by differing dust 
contamination of the primary mirror
(see e.g. \citealt{michard}) over different observing runs are thus insignificant.

\begin{figure*}
\vspace{-20pt}
\includegraphics[width=110mm, angle = -90]{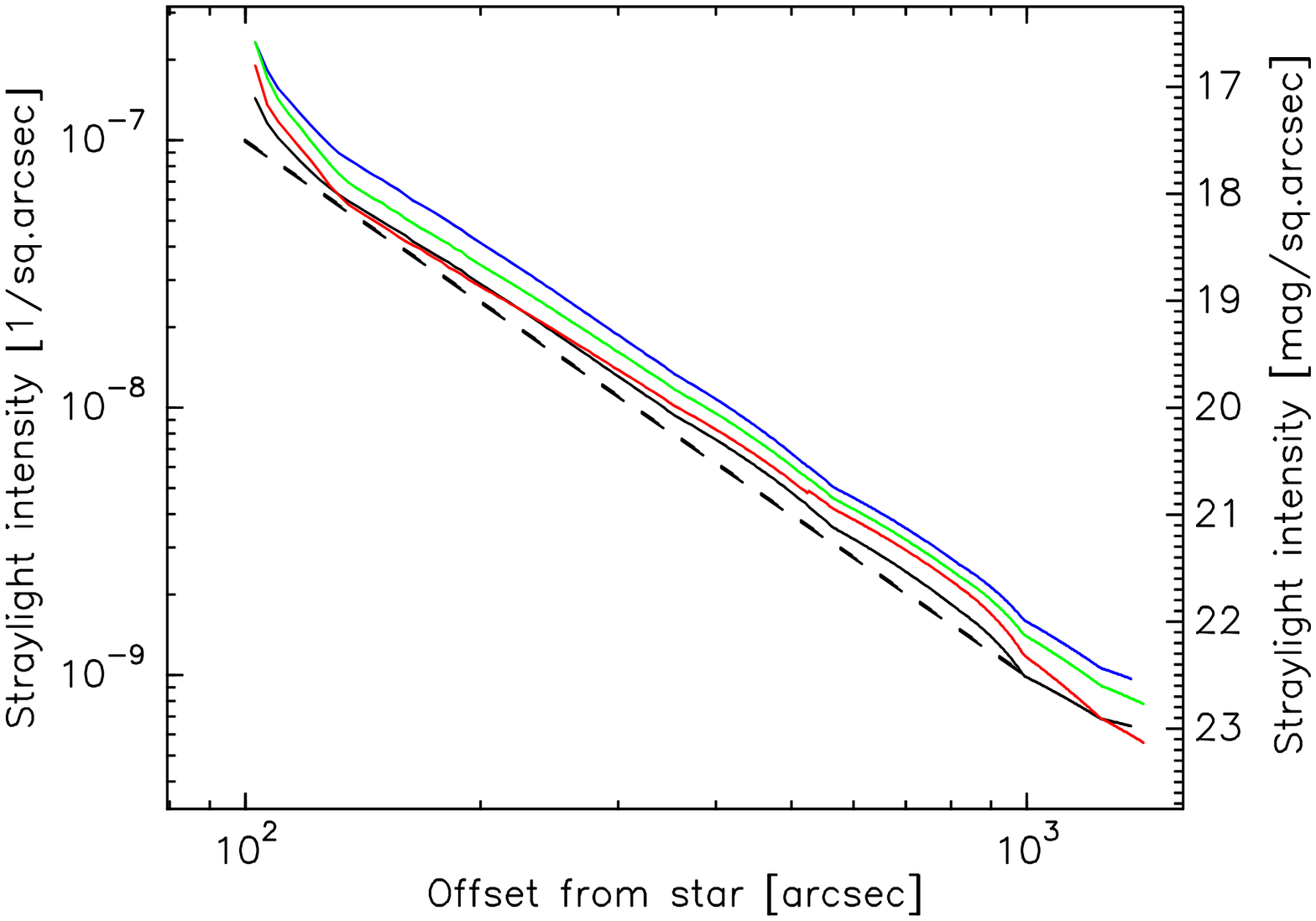}
\vspace{0pt}   
\caption{Star image (aureole) profiles as measured over the offset range of 100\arcsec\, -1400\arcsec\,.
The intensity 
scale on the left-hand y-axis is in units of the star's flux per arcsec$^{2}$ . The
righthand scale is in units of mag\,arcsec$^{-2}$ when the star is of magnitude zero.
Selected wavelength slots are displayed as black (360--380~nm), blue (400--420~nm), green (480--500~nm) 
and red (560--580~nm) lines. For reference the star profile of \citet{King} is shown as dashed line.}
\end{figure*}

\begin{figure*}
\vspace{0pt}
\includegraphics[width=80mm, angle = -90]{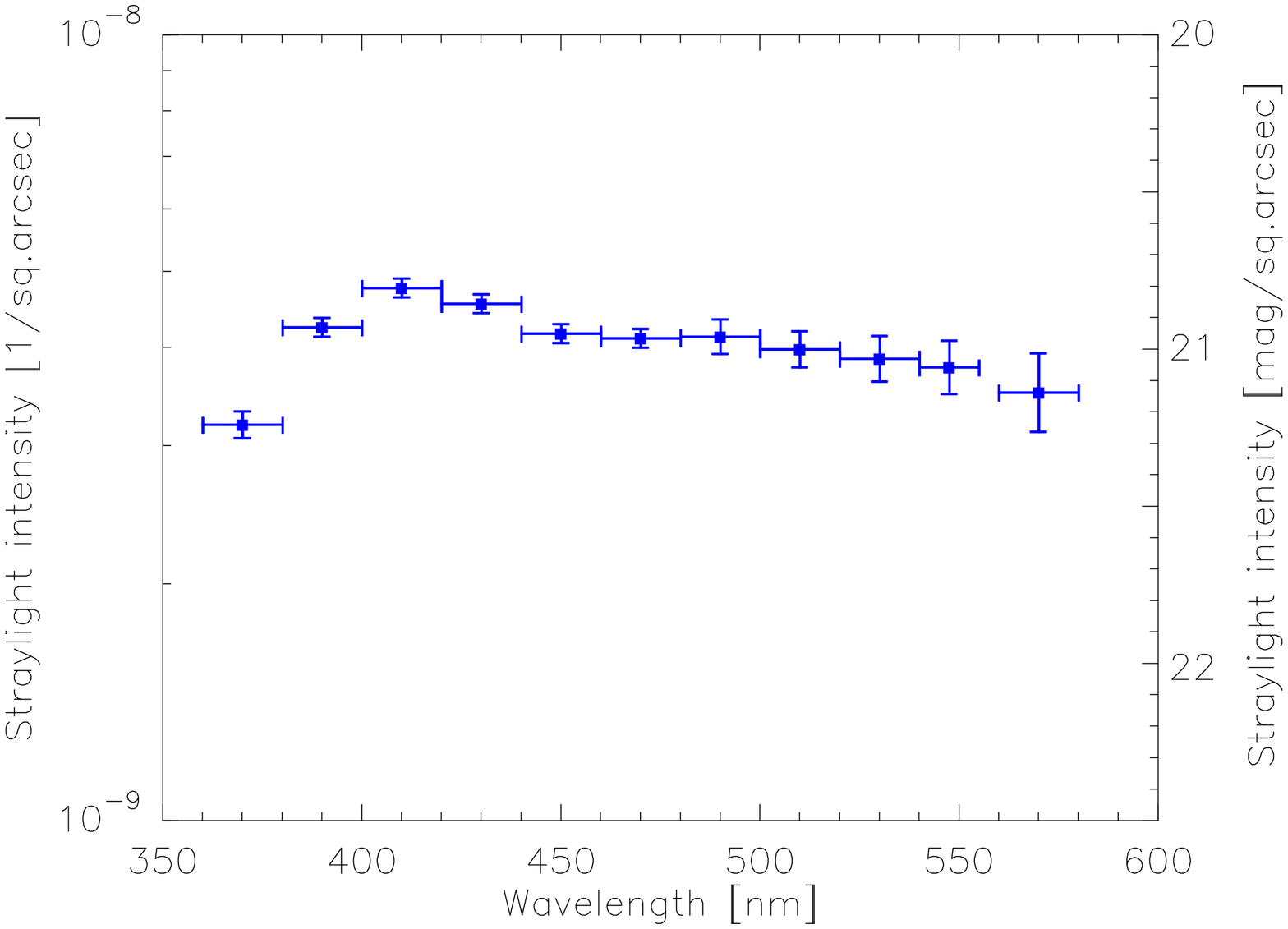}
\caption{Stray light intensity at 600\arcsec\, offset as function of wavelength, calculated from linear
least-squares fits to observed curves between 120\arcsec\,-- 1400\arcsec. Left- and right-hand
y-axis scales are as in Fig. B1. Horizontal bars indicate the wavelength ranges and
the intensity errors are from the fits. }
\end{figure*}

\section{Optical intermediate band and infrared 200 $\bmath{\mu}{\lowercase{\bf m}}$ surface photometry of L\,1642}

An intermediate band surface photometry of selected positions in the area of L\,1642 has 
been carried out in five filters, centred at 360~nm $(u)$, 385~nm, 415~nm, 470~nm $(b)$ and 
555~nm $(y)$, using the ESO 1-m and 50-cm telescopes at La Silla (\citealt{mat90,mat96}; 
see also \citealt{lau}). The 50-cm telescope monitored the airglow variations. 
The observations were made differentially, relative to a standard position (Pos8, see Table 2) 
in the centre of the cloud. Subsequently, the zero level was set by fitting in  Dec, RA 
coordinates a plane through the darkest positions well outside the bright cloud area.  

The ISOPHOT instrument \citep{lemke} aboard the $ISO$ satellite  \citep{kessler} has been used 
to map an area of $\sim$1.2~$\sq\degr$ around L\,1642 at 200~$\mu$m 
\citep{Lehtinen04,Lehtinen07}. In addition, the positions  shown in Fig. 3 as circles 
(photometry only) and squares (photometry and spectrophotometry) were observed also in the 
{\em absolute photometry mode} with ISOPHOT at 200~$\mu$m.

\subsection{Relationship between optical and 200 $\bmath{\mu}{\bf m}$ surface brightness}

The relationship between optical and 200~$\mu$m surface brightness is used in 
 Appendix B to estimate the $B$ and $V$ band
surface brightness distribution in L\,1642. In Fig. C1 the 
optical surface brightness values, $I_{\lambda}^{\rm opt}$, at 415, 470 and 555~nm 
are shown as function of the 200~$\mu$m surface brightness, $I_{200}$.

At small optical depths with $A_{\lambda}\la1$~mag, 
corresponding to  $I_{200}\la20$~MJy,
the relationship is closely linear. At intermediate opacity positions, with 
$A_{\lambda}\sim$1--2~mag, the scattered light has its maximum value. 
For still larger optical depths the scattered light intensity
decreases because of extinction and multiple scattering and absorption losses. 

We have derived a relationship of the following form between the optical and
the 200~$\mu$m surface brightnesses:

\begin{figure} %Figure 9
\vspace{0pt} 
\includegraphics[width=80mm, angle = 0]{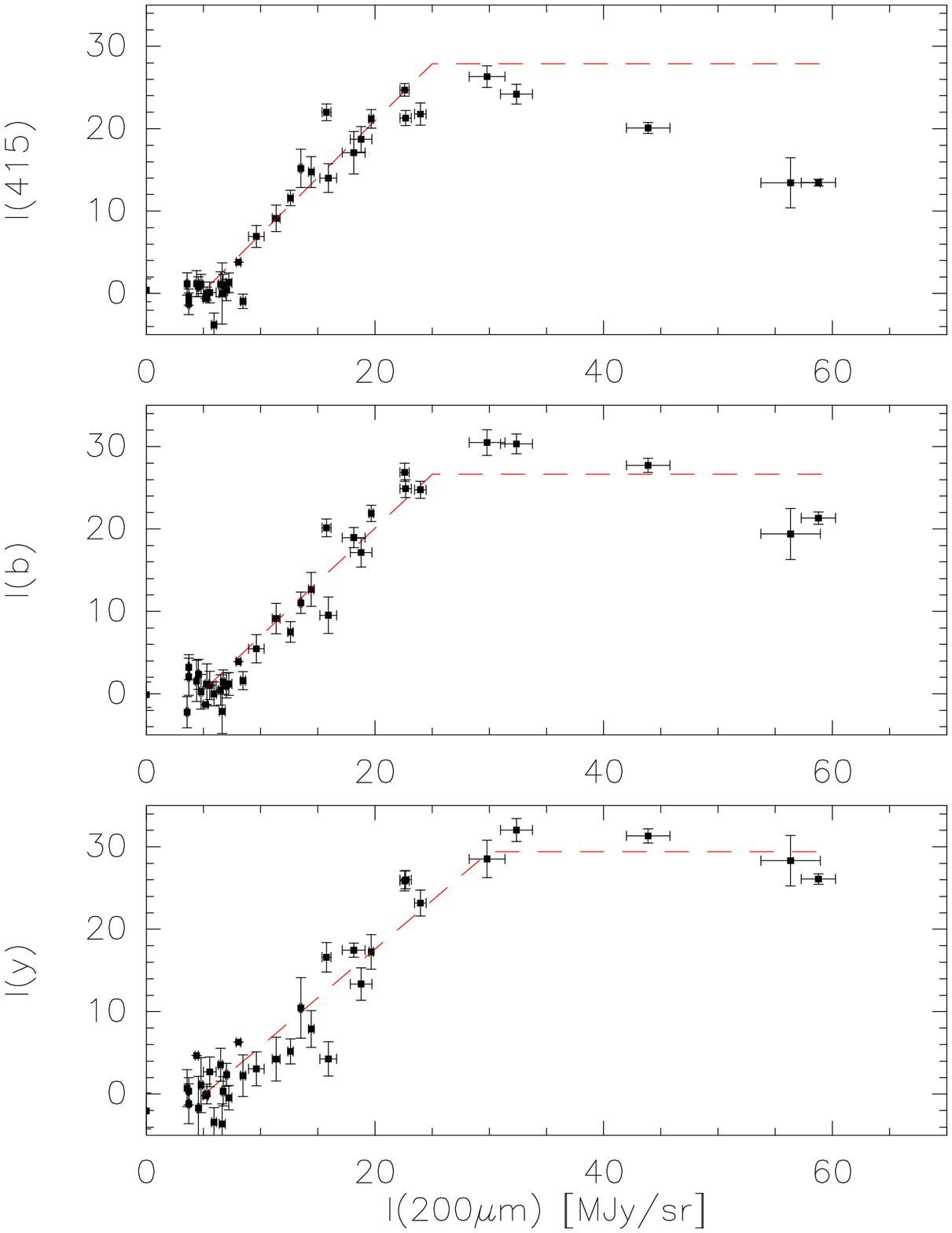}
\caption{Optical surface brightness in the intermediate band filters at  415~nm, 470~nm $(b)$ and 555~nm $(y)$ 
 vs $I_{\rm 200}$ in the L\,1642 area. The unit for optical surface brightness is \cgs\,. } 
%\label{IOPTvsI200}
\end{figure}

\begin{equation}
I_{\lambda}^{\rm opt} = k \cdot I_{200} + b 
\end{equation}
 when $I_{200} \le I_{200}^0$, and
\begin{equation}
I_{\lambda}^{\rm opt} = const. =   k \cdot I^0_{200} + b 
\end{equation}
when $I_{200} >  I^0_{200}$.
Here $I_{200}^0$ designates the value at which the optical surface brightness saturates and the 
linear relationship is no longer valid. Because the high opacity area with 
$I_{200} > I_{200}^0$ covers only a small central part of the cloud 
the approximation with  $I_{\lambda}^{\rm opt}$ = const was considered to be sufficiently good
for the present purpose.
The optical surface brightness values are background-subtracted ones, the background
being defined by the OFF-source positions with the lowest  $I_{200}$ values.
The parameter values for the three observed intermediate bands and the thereof estimated 
values for the $B$ and $V$ bands are given in Table C1.

\hspace{-1cm}
\begin{table}
 \caption{Parameters of the relationship between the optical and the 200~$\mu$m surface brightness 
in the L\,1642 area. The parameters $k$ , $b$ and  $I_{200}^0$ are defined by equations (C1) and (C2). 
The units are $k$: \cgs\,/MJy sr$^{-1}$; $b$: \cgs\, and $I^0_{200}$: MJy\,sr$^{-1}$. }
\begin{tabular}{lccccc}
$\lambda$            &415~nm     & 470~nm   & 550~nm  & $B$      &  $V$\\
\hline
$k$                  &$1.38$     & $1.31$   & $1.18$  &1.35      & 1.18\\
                     &  $\pm0.09$&$\pm0.07$ &$\pm0.07$&$\pm0.09$ &$\pm0.07$ \\
$b$                  & $-6.6$    & $-6.1$   & $-6.0$  &-6.35     &-6.0\\
                     & $\pm0.9$  & $\pm0.9$ & $\pm1.0$& $\pm0.9$ & $\pm1.0$ \\
$I^0_{200}$ & 25        & 25       & 30      & 25       & 30\\
\hline
\end{tabular}
\end{table}

\subsection{Relationship between optical extinction and 200 $\bmath{\mu}{\bf m}$ surface brightness}

For low to moderate extinctions the 200~$\mu$m intensity is linearly correlated with the 
optical extinction,  $A_{\rm V}$, (see e.g. \citealt{Lehtinen07}).
We first used the {\sc nicer} method \citep{lombardi} to derive extinctions from the 2MASS $JHK$ 
colour excesses of stars in $\O$$6\arcmin$ areas corresponding to our spectrophotometry 
positions in Table 1. However, because of the relatively low surface density of 2MASS stars at the high 
galactic latitude of L\,1642 these extinction values had large statistical errors of  
ca. $\pm0.2-0.3$~mag. The high precision achieved for $I_{200}$ 
values (ca. $\pm0.5$~MJy\,sr$^{-1}$) allows a better accuracy  to be reached for $A_{\rm V}$,
especially at low extinctions. 
Therefore, visual extinctions for our low-to-moderate extinction positions, $A_V\la1$mag, 
were determined in the following way using ISOPHOT 200~$\mu$m absolute photometry observations.

A fit to  $I_{200}$ vs $A_{\rm V}$ at $I_{200}<$30 MJy\,sr$^{-1}$ 
gave the slope of $19.0\pm2.5$~MJy\,sr$^{-1}$mag$^{-1}$. 
The zero point of $I_{200}$ was corrected for a
Zodiacal Emission (ZE) of $0.8\pm0.2$~MJy\,sr$^{-1}$ and a CIB of $1.1\pm0.3$~MJy\,sr$^{-1}$ \citep{hau}. 
The ZE at time of our 200~$\mu$m observations (1998-03-19/20, 
longitude difference $\lambda$(L\,1642)-$\lambda_{\sun}=63\fdg 7)$
was estimated using a 270 K blackbody fit to the ZE intensities at 100, 140 and 
240~$\mu$m. They were interpolated from the weekly DIRBE Sky and Zodi Atlas (DSZA)
\footnote{https://lambda.gsfc.nasa.gov/product/.../dirbe\_dsza\_data\_get.cfm}
maps based on the \citet{kelsall} interplanetary dust distribution model. 
The extinction values were then calculated from \\
$A_{\rm V} = (I_{200} - 1.9$~MJy\,sr$^{-1}$)/19~MJy\,sr$^{-1}$mag$^{-1}$ \\
and are given for our spectrophotometric positions in Table~1.
We estimate their errors to be ca. $\pm0.05$~mag.

\section{The spectra in numerical form}
{ The resulting spectra for Positions 8, 9, 42, and the mean of 9 and 42 as discussed 
in Section 8 and displayed in Fig.~6 are available in the form of machine readable ASCII.txt 
files as an attachment to the electronic version of this Paper. The files are named 
Pos8.txt, Pos9.txt, Pos42.txt, and Pos9\_42.txt. Each file has two columns: the wavelength
ranging from 3649 to 5955 \AA\, and the intensity in units of \cgs\,.}


\begin{thebibliography}{99}

\bibitem[\protect\citeauthoryear{Aharonian et al.}{2006}]{Ah} Aharonian, F., et al., 2006, Nature, 440, 1018
\bibitem[\protect\citeauthoryear{Ackermann et. al.}{2012}]{Ackermann} Ackermann, M.,et al., 2012, Science, 338, 1190
\bibitem[\protect\citeauthoryear{Appenzeller et al.}{1998}]{app} Appenzeller I., et al., 1998, Msngr, 94, 1 
\bibitem[\protect\citeauthoryear{Ashburn}{1954}]{Ashburn} Ashburn E.~V., 1954, JATP 5, 83 (reprinted in JGR, 59, 67) 
\bibitem[\protect\citeauthoryear{Beckers}{1995}]{Beckers} Beckers J.~M., 1995, in International Symposium on 
the Scientific and Engineering Frontiers for 8 - 10 m Telescopes, p. 303 
\bibitem[\protect\citeauthoryear{Bernstein}{2007}]{b7} Bernstein R.~A., 2007, ApJ, 666, 663 
\bibitem[\protect\citeauthoryear{Bernstein et al.}{2002a}]{b1} Bernstein, R.A., Freedman, W.~L., \& Madore, B.~F. 2002 ApJ, 571, 56
\bibitem[\protect\citeauthoryear{Bernstein et al.}{2002b}]{b2} Bernstein, R.A. et al. 2002 ApJ, 571, 107
\bibitem[\protect\citeauthoryear{Bernstein et al.}{2005}]{b5} Bernstein, R.A., Freedman, W.~L., \& Madore, B.~F.  2005 ApJ, 632, 713
\bibitem[\protect\citeauthoryear{Biteau \& Williams}{2015}]{biteau} Biteau J., Williams D.~A., 2015, ApJ, 812, 60
\bibitem[\protect\citeauthoryear{Bruzual A.}{1983}]{bruzual} Bruzual A.~G., 1983, ApJ, 273, 105 

\bibitem[\protect\citeauthoryear{Chandrasekhar}{1950}]{chandra} Chandrasekhar, S., 1950, Radiative Transfer. Oxford, Clarendon Press 
\bibitem[\protect\citeauthoryear{Dom{\'{\i}}nguez et al.}{2013}]{Dom}  Dom{\'{\i}}nguez A., Finke J.~D., Prada F., Primack J.~R., 
Kitaura F.~S., Siana B., Paneque D., 2013, ApJ, 770, 77
\bibitem[\protect\citeauthoryear{Dressler \& Shectman}{1987}]{dressler} Dressler A., Shectman S.~A., 1987, AJ, 94, 899 

\bibitem[\protect\citeauthoryear{Hamilton}{1985}]{hamilton} Hamilton D., 1985, ApJ, 297, 371 
\bibitem[\protect\citeauthoryear{Hauser et al.}{1998}]{hau} Hauser M.~G., et al., 1998, ApJ, 508, 25
\bibitem[\protect\citeauthoryear{Hearty et al.}{2000}]{hearty} 
Hearty T., Fern{\'a}ndez M., Alcal{\'a} J.~M., Covino E., Neuh{\"a}user R., 2000, A\&A, 357, 681 
\bibitem[\protect\citeauthoryear{H.E.S.S. Collaboration}{2013}]{hess} H.E.S.S. Collaboration 2013, A\&A 550, A4
\bibitem[\protect\citeauthoryear{H$\o$g et al.}{2000}]{Hog} H$\o$g, E., et al., 2000, A\&A, 355, L27 
\bibitem[\protect\citeauthoryear{Juvela et al.}{2009}]{juvela} Juvela M., Mattila K., Lemke D., Klaas U., Leinert C., Kiss C., 2009, A\&A, 500, 763 
\bibitem[\protect\citeauthoryear{Kelsall et al.}{1998}]{kelsall} Kelsall, T. et al. 1998 ApJ, 508, 44
\bibitem[\protect\citeauthoryear{Kessler et al.}{1996}]{kessler} Kessler M.~F., et al., 1996, A\&A, 315, L27 
\bibitem[\protect\citeauthoryear{King}{1971}]{King} King I.~R., 1971, PASP, 83, 199 
\bibitem[\protect\citeauthoryear{Kurucz et al.}{1984}]{Kur} 
Kurucz R.~L., Furenlid I., Brault J., Testerman L., 1984, Solar Flux Atlas from 296 to 1300~nm 
\bibitem[\protect\citeauthoryear{Laureijs, Mattila, \& Schnur}{1987}]{lau} 
Laureijs R.~J., Mattila K., Schnur G., 1987, A\&A, 184, 269
\bibitem[\protect\citeauthoryear{Lehtinen et al.}{2004}]{Lehtinen04} 
Lehtinen K., Russeil D., Juvela M., Mattila K., Lemke D., 2004, A\&A, 423, 975 
\bibitem[\protect\citeauthoryear{Lehtinen et al.}{2007}]{Lehtinen07} Lehtinen K., Juvela M., Mattila K., Lemke D., Russeil D., 2007, A\&A, 466, 969
\bibitem[\protect\citeauthoryear{Lehtinen \& Mattila}{2013}]{Lehtinen13} Lehtinen K., Mattila K., 2013, A\&A, 549, A91 
 Lehtinen K., Juvela M., Mattila K., Lemke D., Russeil D., 2007, A\&A, 466, 969 
\bibitem[\protect\citeauthoryear{Leinert et al.}{1998}]{Leinert} Leinert, C. et al. 1998, A\&AS 127, 1 
\bibitem[\protect\citeauthoryear{Lemke et al.}{1996}]{lemke} Lemke D., et al., 1996, A\&A, 315, L64 
\bibitem[\protect\citeauthoryear{Lombardi \& Alves}{2001}]{lombardi} Lombardi M., Alves J., 2001, A\&A, 377, 1023 
\bibitem[\protect\citeauthoryear{Longair}{1995}]{longair95} Longair M.~S., 1995,
in The Deep Universe, ed. B.~Binggeli \& R.~ Buser, (Saas-Fee Advanced Course 23; Berlin: Springer), 317
\bibitem[\protect\citeauthoryear{Matsuoka et al.}{2011}]{Matsu11} Matsuoka, Y., Ienaka, N., Kawara, K.,
 Oyabu S., 2011, ApJ, 736, 119 
\bibitem[\protect\citeauthoryear{Mattila}{1970}]{mat70} {Mattila, K. 1970, A\&A, 9, 53}
\bibitem[\protect\citeauthoryear{Mattila}{1976}]{mat76} {Mattila, K. 1976, A\&A, 47, 77}
\bibitem[\protect\citeauthoryear{Mattila}{1980a}]{mat80a} Mattila K., 1980a, A\&A, 82, 373
\bibitem[\protect\citeauthoryear{Mattila}{1980b}]{mat80b} Mattila K., 1980b, A\&AS, 39, 53 
\bibitem[\protect\citeauthoryear{Mattila \& Schnur}{1990}]{mat90} {Mattila, K. \& Schnur,G. 1990, IAU Symposium No. 139, 257}
\bibitem[\protect\citeauthoryear{Mattila, V\"ais\"anen, \& von Appen-Schnur}{1996}]{mat96} 
Mattila K.,  V\"ais\"anen P., von Appen-Schnur G.~F.~O., 1996, A\&AS, 119, 153 
\bibitem[\protect\citeauthoryear{Mattila}{2003}]{mat03} {Mattila, K. 2003, ApJ, 591, 119}
\bibitem[\protect\citeauthoryear{Mattila et al.}{2012}]{mat12} 
Mattila K., Lehtinen K., V{\"a}is{\"a}nen P., von Appen-Schnur G., Leinert C., 2012, IAUS, 284, 429 
\bibitem[\protect\citeauthoryear{Mattila et al.}{2017b}]{mat17b} Mattila K., V{\"a}is{\"a}nen P.,  Lehtinen K., 
von Appen-Schnur G., Leinert C., 2017b,  MNRAS, xx, yy (Paper~II) 
\bibitem[\protect\citeauthoryear{Michard}{2002}]{michard} Michard R., 2002, A\&A, 384, 763
\bibitem[\protect\citeauthoryear{Partridge \& Peebles}{1967}]{P67} Partridge, B., Peebles, P.J.E. 1967, ApJ, 148, 377
\bibitem[\protect\citeauthoryear{Patat}{2003}]{Patat03} Patat F., 2003, A\&A, 400, 1183 
\bibitem[\protect\citeauthoryear{Patat et al.}{2011}]{Patat11} Patat F., et al., 2011, A\&A, 527, A91 
\bibitem[\protect\citeauthoryear{Puget et al.}{1996}]{pug} 
Puget J.-L., Abergel A., Bernard J.-P., Boulanger F., Burton W.~B., Desert F.-X., Hartmann D., 1996, A\&A, 308, L5 
\bibitem[\protect\citeauthoryear{Tody}{1993}]{Tody} Tody D., 1993, ASPC, 52, 173
\bibitem[\protect\citeauthoryear{Toller}{1983}]{Tol} Toller G.~N., 1983, ApJ, 266, L79
\bibitem[\protect\citeauthoryear{T{\"u}g}{1977}]{Tug} T{\"u}g H., 1977, Msngr, 11, 7 
\bibitem[\protect\citeauthoryear{V\"ais\"anen}{1996}]{Vai}{V\"ais\"anen,P. 1996, A\&A 315, 21}
\bibitem[\protect\citeauthoryear{Windhorst et al.}{2011}]{Windhorst} Windhorst R.~A., et al., 2011, ApJS, 193, 27
\bibitem[\protect\citeauthoryear{Zacharias et al.}{2004}]{Nomad04} Zacharias N., Monet D.~G., 
Levine S.~E., Urban S.~E., Gaume R., Wycoff G.~L., 2004, AAS, 36, 1418 
\bibitem[\protect\citeauthoryear{Zacharias et al.}{2005}]{Nomad05} Zacharias N., Monet D.~G., Levine S.~E., 
Urban S.~E., Gaume R., Wycoff G.~L., 2005, yCat, 1297, 0 
\bibitem[\protect\citeauthoryear{Zemcov et al.}{2014}]{Zem} Zemcov M., et al., 2014, Sci, 346, 732 
\bibitem[\protect\citeauthoryear{Zemcov et al.}{2017}]{zemcov17} Zemcov M., Immel P., Nguyen C., Cooray A., Lisse C.~M., Poppe A.~R., 2017, Nature Communications, 8, A15003 


\end{thebibliography}
\end{document}